\documentclass[sigconf]{acmart}

\usepackage{graphicx, subcaption}
\usepackage{multirow}
\usepackage[normalem]{ulem}
\useunder{\uline}{\ul}{}
\usepackage{enumitem}
\usepackage{color}

\usepackage[noend]{algpseudocode}
\usepackage[ruled]{algorithm2e}
\usepackage{hyperref}
\hypersetup{colorlinks=false,linkcolor=blue,urlcolor=blue,citecolor=red}
\usepackage{balance}

\newcommand\norm[1]{\left\lVert#1\right\rVert}
\newcommand{\Lapl}{\mathbf{\mathop{\mathcal{L}}}}

\newcommand{\Trans}[1]{{#1}^{\top}}

\newcommand{\Mat}[1]{\textbf{#1}}

\newcommand{\Set}[1]{\mathcal{#1}}

\newcommand{\ie}{\emph{i.e., }}
\newcommand{\eg}{\emph{e.g., }}

\newcommand{\wrt}{\emph{w.r.t. }}
\newcommand{\cf}{\emph{cf. }}

\AtBeginDocument{%
  \providecommand\BibTeX{{%
    \normalfont B\kern-0.5em{\scshape i\kern-0.25em b}\kern-0.8em\TeX}}}

\copyrightyear{2021}
\acmYear{2021}
\setcopyright{acmcopyright}\acmConference[SIGIR '21]{Proceedings of the 44th
International ACM SIGIR Conference on Research and Development in Information
Retrieval}{July 11--15, 2021}{Virtual Event, Canada}
\acmBooktitle{Proceedings of the 44th International ACM SIGIR Conference on
Research and Development in Information Retrieval (SIGIR '21), July 11--15, 2021,
Virtual Event, Canada}
\acmPrice{15.00}
\acmDOI{10.1145/3404835.3462862}
\acmISBN{978-1-4503-8037-9/21/07}

\begin{document}
\fancyhead{}

\title{Self-supervised Graph Learning for Recommendation}






\author{Jiancan Wu$^1$, Xiang Wang$^{2*}$, Fuli Feng$^2$, Xiangnan He$^1$, Liang Chen$^3$, Jianxun Lian$^4$, and Xing Xie$^4$}

\def \authors{Jiancan Wu, Xiang Wang, Fuli Feng, Xiangnan He, Liang Chen, Jianxun Lian, and Xing Xie}

\affiliation{
\institution{$^1$University of Science and Technology of China \country{China}}
}
\affiliation{
\institution{$^2$National University of Singapore \country{Singapore}}
}
\affiliation{
\institution{$^3$Sun Yat-sen University \country{China},
$^4$Microsoft Research Asia \country{China}}
}

\email{wjc1994@mail.ustc.edu.cn, {xiangwang1223,fulifeng93,xiangnanhe}@gmail.com}
\email{chenliang6@mail.sysu.edu.cn,{jianxun.lian,xing.xie}@microsoft.com}

\thanks{$*$Xiang Wang is the corresponding author}


\begin{abstract}
	Representation learning on user-item graph for recommendation has evolved from using single ID or interaction history to exploiting higher-order neighbors. This leads to the success of graph convolution networks (GCNs) for recommendation such as PinSage and LightGCN. Despite effectiveness, we argue that they suffer from two limitations: (1) high-degree nodes exert larger impact on the representation learning, deteriorating the recommendations of low-degree (long-tail) items; and (2) representations are vulnerable to noisy interactions, as the neighborhood aggregation scheme further enlarges the impact of observed edges.

	In this work, we explore self-supervised learning on user-item graph, so as to improve the accuracy and robustness of GCNs for recommendation.
	The idea is to supplement the classical supervised task of recommendation with an auxiliary self-supervised task, which reinforces node representation learning via self-discrimination. 
	Specifically, we generate multiple views of a node, maximizing the agreement between different views of the same node compared to that of other nodes. We devise three operators to generate the views --- 
    node dropout, edge dropout, and random walk
	--- that change the graph structure in different manners.
	We term this new learning paradigm as \textit{Self-supervised Graph Learning} (SGL), implementing it on the state-of-the-art model LightGCN.
	Through theoretical analyses, we find that SGL has the ability of automatically mining hard negatives. 
	Empirical studies on three benchmark datasets demonstrate the effectiveness of SGL, which improves the recommendation accuracy, especially on long-tail items, and  the robustness against interaction noises.
    Our implementations are available at \url{https://github.com/wujcan/SGL}.
\end{abstract}



\maketitle

\section{Introduction}
Learning high-quality user and item representations from interaction data is the theme of collaborative recommendation. 
Earlier work like matrix factorization (MF)~\cite{BPR} projects single ID of each user (or item) into an embedding vector.
Some follow-on studies~\cite{SVD++,NAIS} 
enrich the single ID with interaction history for learning better representations. 
More recently, representation learning has evolved to exploiting higher-order connectivity in user-item graph.
The technique is inspired from the graph convolution networks (GCNs), which provide an end-to-end way to integrate multi-hop neighbors into node representation learning and achieve state-of-the-art performance for recommendation~\cite{NGCF,LightGCN,GCMC,PinSage}.

Despite effectiveness, 
current GCN-based recommender models suffer from some limitations:
\begin{itemize}[leftmargin=*]
    \item \textbf{Sparse Supervision Signal.} Most models approach the recommendation task under a supervised learning paradigm~\cite{BPR,NeuMF,LightGCN}, where the supervision signal comes  from the observed user-item interactions.
    However, the observed interactions are extremely sparse~\cite{DBLP:conf/www/HeM16,iCD} compared to the whole interaction space, making it  insufficient to learn quality representations. 
    
    \item \textbf{Skewed Data Distribution.} Observed interactions usually exhibit a power-law distribution~\cite{DBLP:journals/siamrev/ClausetSN09,PowerLaw}, where 
    the long tail consists of low-degree items that lack supervision signal.
    In contrast, high-degree items 
    appear more frequently in neighborhood aggregation and supervised loss, thus exert larger impact on representation learning. 
    Hence, the GCNs are easily biased towards high-degree items~\cite{tang2020investigating,abs-2010-03240}, sacrificing the performance of low-degree (long-tail) items.

    \item \textbf{Noises in Interactions.} Most feedback that a user provides is implicit (\eg clicks, views), instead of explicit (\eg ratings, likes/dislikes). As such, observed interactions usually contain noises, \eg a user is misled to click an item and finds it uninteresting after consuming it~\cite{abs-2006-04153}.
    The neighborhood aggregation scheme in GCNs enlarges the impact of interactions on representation learning, making the learning more vulnerable to interaction noises.
\end{itemize}

In this work, we focus on exploring self-supervised learning (SSL) in recommendation, to solve the foregoing limitations.
Though being prevalent in computer vision (CV)~\cite{DBLP:conf/iclr/GidarisSK18,CPC} and natural language processing (NLP)~\cite{DBLP:conf/naacl/DevlinCLT19,DBLP:conf/iclr/LanCGGSS20}, SSL is relatively less explored in recommendation. The idea is to set an auxiliary task that distills additional signal from the input data itself, especially through exploiting the unlabeled data space. For example, BERT~\cite{DBLP:conf/naacl/DevlinCLT19} randomly masks some tokens in a sentence, setting the prediction of the masked tokens as the auxiliary task that can capture the dependencies among tokens; RotNet~\cite{DBLP:conf/iclr/GidarisSK18} randomly rotates labeled images, training the model on the rotated images to get improved representations for the mask task of object recognition or image classification.
Compared with supervised learning, SSL allows us 
to exploit the unlabeled data space via making changes on the input data, achieving 
remarkable improvements in downstream tasks~\cite{SimCLR}.

Here we wish to bring the SSL's superiority into recommendation representation learning, which differs from CV/NLP tasks since the data are discrete and inter-connected. To address the aforementioned limitations of GCN-based recommendation models, we construct the auxiliary task as discriminating the representation of a node itself. Specifically, it consists of two key components: 
(1) \textbf{data augmentation}, which generates multiple views for each node,
and (2) \textbf{contrastive learning}, which maximizes the agreement between different views of the same node, compared to that of other nodes.
For GCNs on user-item graph, the graph structure serves as the input data that plays an essential role for representation learning. 
From this view, it is natural to construct the "unlabeled" data space by changing the graph adjacency matrix, and we develop three operators to this end: 
node dropout, edge dropout, and random walk,
where each operator works with a different rationality.
Thereafter, we perform contrastive learning based on the GCNs on the changed structure.
As a result, SGL augments the node representation learning by exploring the internal relationship among nodes.

Conceptually, our SGL 
supplements existing GCN-based recommendation models in:
(1) node self-discrimination offers auxiliary supervision signal, which is complementary to the classical supervisions 
from observed interactions only;
(2) the augmentation operators, especially edge dropout, helps to mitigate the degree biases by intentionally reducing the influence of high-degree nodes;
(3) the multiple views for nodes \wrt different local structures and neighborhoods enhance the model robustness against interaction noises.
Last but not least, we offer theoretical analyses for the contrastive learning paradigm, finding that it has the side effect of mining hard negative examples, which not only boosts the performance but also accelerates the training process. 

It is worthwhile mentioning that our SGL is model-agnostic and can be applied 
to any graph-based model that consists of user and/or item embedding. 
Here we implement it on the simple but effective model, LightGCN~\cite{LightGCN}.
Experimental studies on three benchmark datasets demonstrate the effectiveness of SGL, which significantly improves the recommendation accuracy, especially on long-tail items, and enhance the robustness against interaction noises.
We summarize the contributions of this work as follows:
\begin{itemize}[leftmargin=*]
    \item We devise a new learning paradigm, SGL, which takes node self-discrimination as the self-supervised task to offer auxiliary signal for representation learning. 
    \item In addition to mitigating degree bias and increasing robustness to interaction noises, we prove in theory that SGL inherently encourages learning from hard negatives, controlled by the temperature hyper-parameter in the softmax loss function.
    \item We conduct extensive experiments on three benchmark datasets to demonstrate the superiority of SGL. 
\end{itemize}

\section{Preliminaries}\label{sec:preliminaries}
We first summarize the common paradigm of GCN-based collaborative filtering models.
Let $\Set{U}$ and $\Set{I}$ be the set of users and items respectively.
Let $\Set{O}^{+}=\{y_{ui}|u\in\Set{U},i\in\Set{I}\}$ be the observed interactions, where $y_{ui}$ indicates that user $u$ has adopted item $i$ before.
Most existing models~\cite{NGCF,LightGCN,GCMC} construct a bipartite graph $\Set{G}=(\Set{V},\Set{E})$, where the node set $\Set{V}=\Set{U}\cup\Set{I}$ involves all users and items, and the edge set $\Set{E}=\Set{O}^{+}$ represents observed interactions.

\noindent\textbf{Recap GCN.}
At the core is to apply the neighborhood aggregation scheme on $\Set{G}$, updating the representation of ego node by aggregating the representations of neighbor nodes:
\begin{gather}
    \Mat{Z}^{(l)}=H(\Mat{Z}^{(l-1)},\Set{G})
\end{gather}
where $\Mat{Z}^{(l)}$ denotes the node representations at the $l$-th layer, $\Mat{Z}^{(l-1)}$ is that of previous layer, and $\Mat{Z}^{(0)}$ is the ID embeddings (trainable parameters). $H$ denotes the function for neighborhood aggregation, which is more interpretable from the vector level:
\begin{equation}
    \Mat{z}^{(l)}_{u}=f_{\text{combine}}(\Mat{z}^{(l-1)}_{u}, f_{\text{aggregate}}(\{\Mat{z}^{(l-1)}_{i}|i\in\Set{N}_{u}\})).
\end{equation}
To update the representation of ego node $u$ at the $l$-th layer, it first aggregates the representations of its neighbors $\mathcal{N}_u$ at the $(l-1)$-th layer, then combines with its own representation $\Mat{z}^{(l-1)}_{u}$.
There are a number of designs for $f_{\text{aggregate}}(\cdot)$ and $f_{\text{combine}}(\cdot)$~\cite{GraphSage,GCN,GAT,GIN}.
We can see that the representations of the $l$-th layer encode the $l$-order neighbors in the graph. After obtaining $L$ layers representations, there may be a readout function to generate the final representations for prediction: 
\begin{gather}
    \Mat{z}_{u}=f_{\text{readout}}\Big(\{\Mat{z}^{(l)}_{u}|l=[0,\cdots,L]\}\Big).
\end{gather}
Common designs include last-layer only~\cite{GCMC,PinSage}, concatenation~\cite{NGCF}, and weighted sum~\cite{LightGCN}. 

\noindent\textbf{Supervised Learning Loss.}
A prediction layer is built upon the final representations to predict how likely $u$ would adopt $i$.
A classical solution is the inner product, which supports fast retrieval:
\begin{gather}
    \hat{y}_{ui}=\Trans{\Mat{z}}_{u}\Mat{z}_{i}.
\end{gather}
To optimize model parameters, existing works usually frame the task as one of supervised learning, where the supervision signal comes from the observed interactions (also the edges of $\mathcal{G}$).
For example, encouraging the predicted value $\hat{y}_{ui}$ to be close to the ground truth value $y_{ui}$ and selecting negative examples from missing data~\cite{NeuMF}. 
Besides above point-wise learning, another common choice is the pairwise Bayesian Personalized Ranking (BPR) loss~\cite{BPR}, which enforces the prediction of an observed interaction to be scored higher than its unobserved counterparts:
\begin{gather}\label{eq:supervised-loss}
    \Lapl_{main}=\sum_{(u,i,j)\in\Set{O}}-\log\sigma(\hat{y}_{ui}-\hat{y}_{uj}),
\end{gather}
where $\Set{O}=\{(u,i,j)|(u,i)\in\Set{O}^{+},(u,j)\in\Set{O}^{-}\}$ is the training data, and $\Set{O}^{-}=\Set{U}\times\Set{I}\setminus\Set{O}^{+}$ is the unobserved interactions. In this work, we choose it as the main supervised task. 
\section{Methodology}
\begin{figure}
    \centering
    \includegraphics[width=0.46\textwidth]{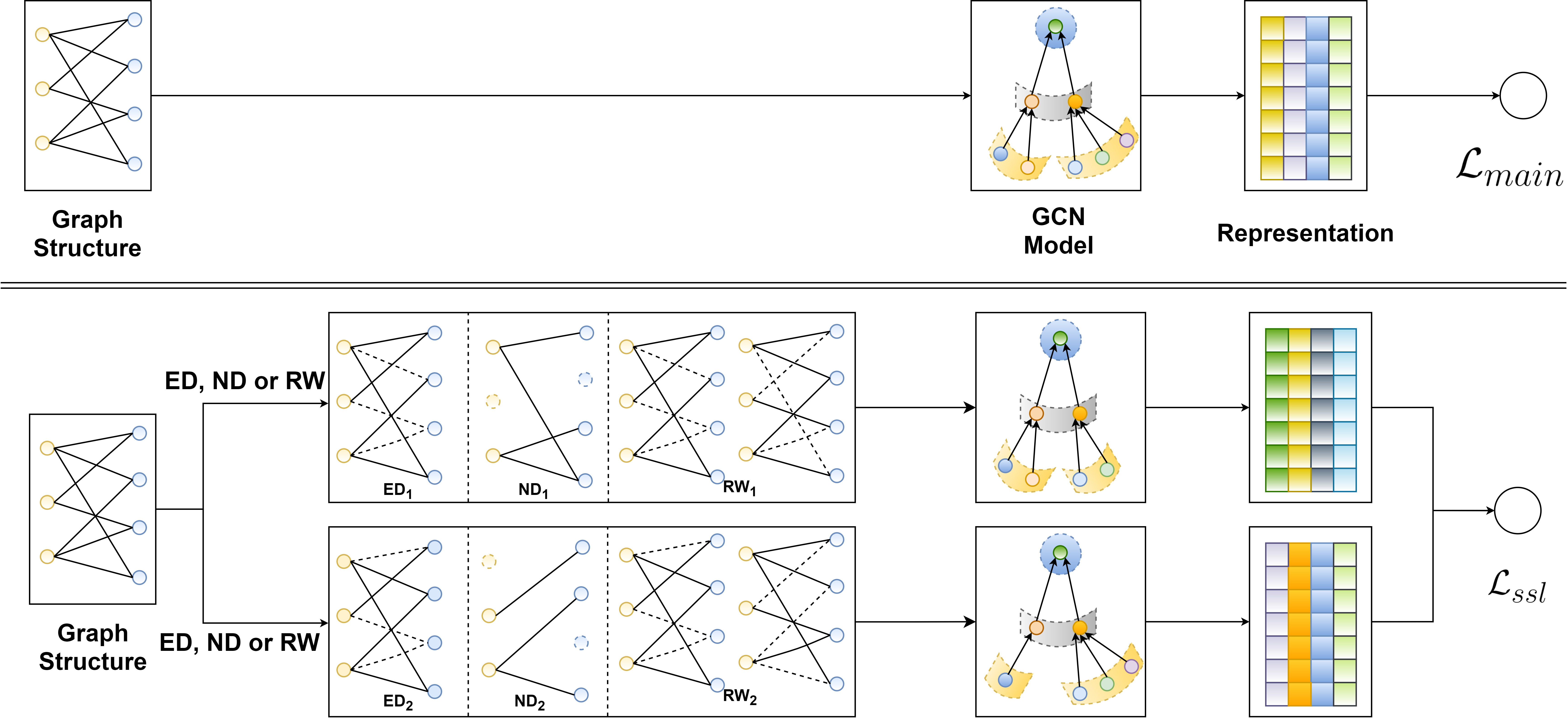}
    \vspace{-10pt}
    \caption{The overall system framework of SGL. The upper layer illustrates the working flow of the main supervised learning task while the bottom layer shows the working flows of SSL task with augmentation on graph structure.}
    \label{fig:system_framework}
    \vspace{-10pt}
\end{figure}

We present the proposed Self-supervised Graph Learning (SGL) paradigm, which supercharges the main supervised task with self-supervised learning. 
Figure~\ref{fig:system_framework} illustrates the working flow of SGL. 
Specifically, the self-supervised task (also termed as pretext task or auxiliary task) is to construct supervision signal from the correlation within the input data.

Specifically, we introduce how to perform data augmentation that generates multiple representation views, followed by the contrastive learning based on the generated representations to build the pretext task. SSL is combined with classical GCN in a multi-task learning manner.
Thereafter, we conduct theoretical analyses on SSL from the gradient level, revealing the connection with hard negative mining.
Lastly, we analyze the complexity of SGL. 

\subsection{Data Augmentation on Graph Structure}\label{sec:data_augmentation}
Directly grafting the data augmentation adopted in CV and NLP tasks~\cite{ContrastiveLearning,SimCLR,MoCo,DBLP:conf/naacl/DevlinCLT19} is infeasible for graph-based recommendation, due to specific characteristics:
(1) 
The features of users and items are discrete, like one-hot ID and other categorical variables. 
Hence, the augmentation operators on images, such as random crop, rotation, or blur, are not applicable.
(2) More importantly, unlike CV and NLP tasks that treat each data instance as isolated, 
users and items in the interaction graph are inherently connected and dependent on each others.
Thus, we need new augmentation operators tailored for graph-based recommendation.

\begin{figure}[t]
	\centering
	\subcaptionbox{Edge Dropout\label{fig:toy_ED}}{
		\includegraphics[width=0.23\textwidth]{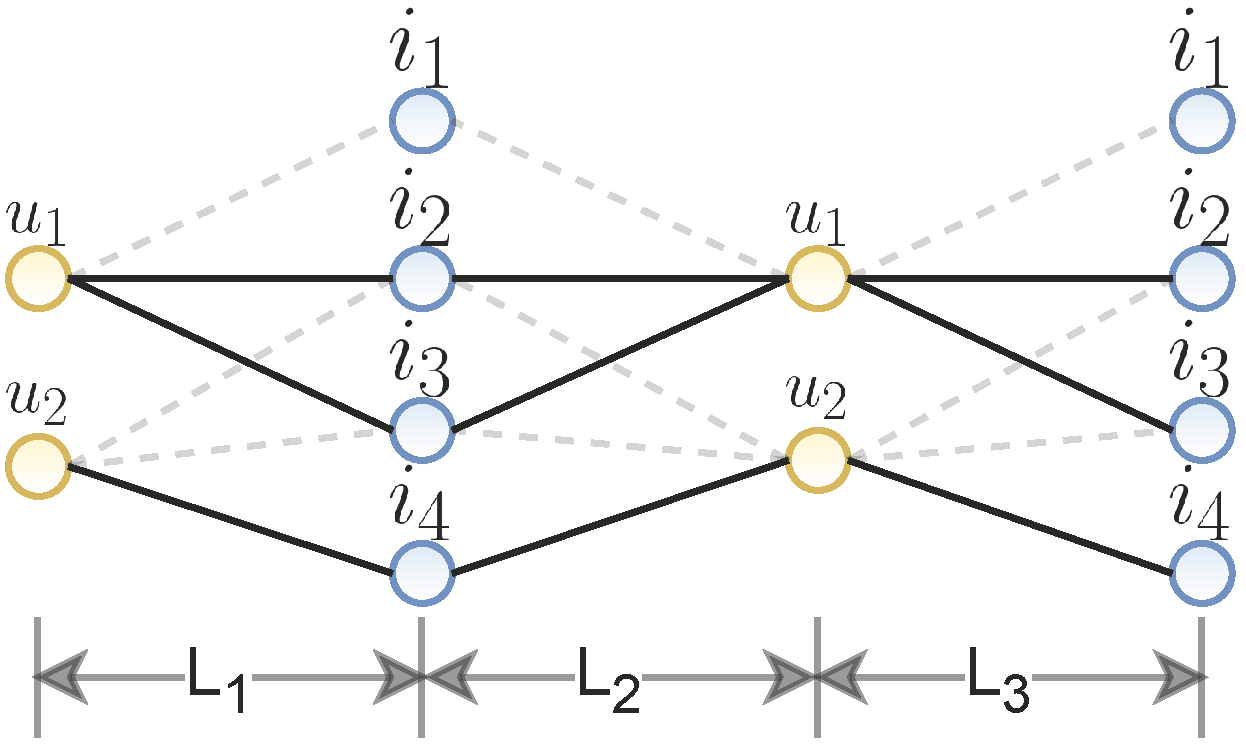}}
	\subcaptionbox{Random Walk\label{fig:toy_RW}}{
		\includegraphics[width=0.23\textwidth]{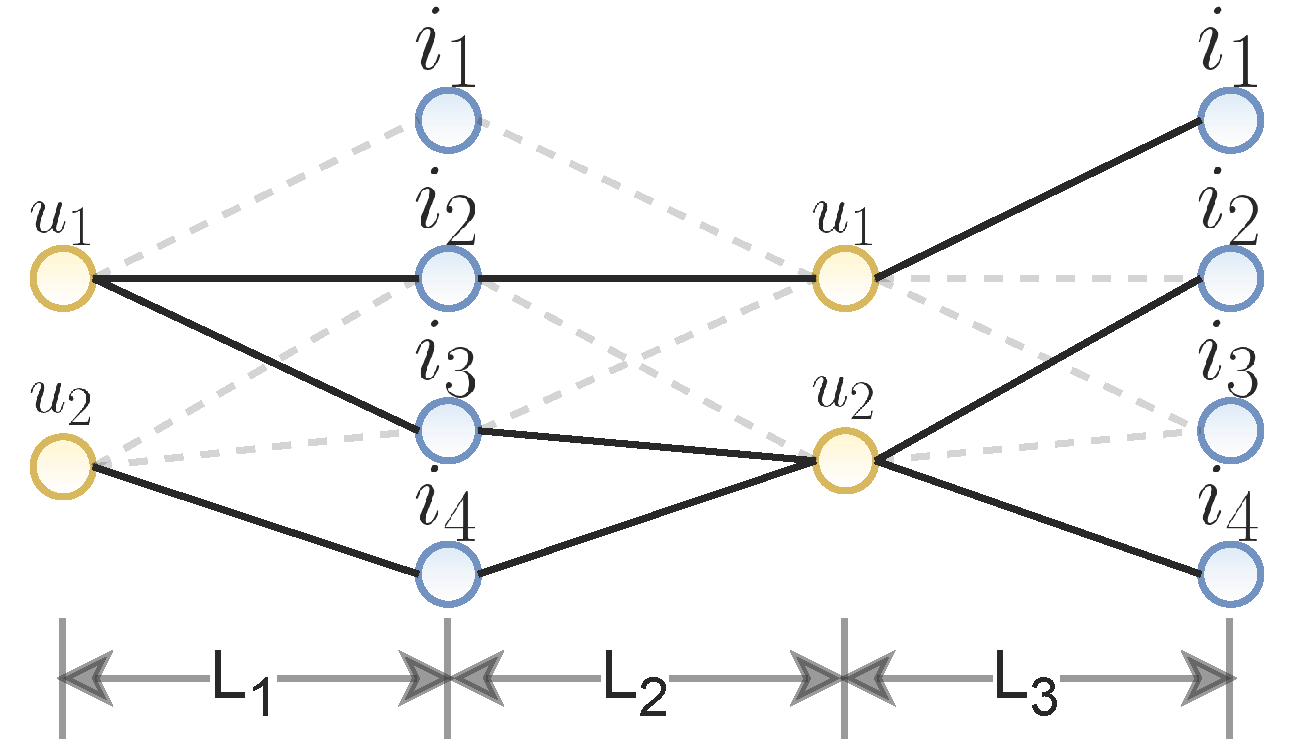}}
	\vspace{-10pt}
	\caption{A toy example of higher-order connectivity in a three-layer GCN model with Edge Dropout (left) and Random Walk (right).
    For Random Walk, the graph structure keeps changing across layers as opposed to Edge Dropout. As a result, there exists a three-order path between node $u_1$ and $i_1$ that does not exist in Edge Dropout.}
	\label{fig:toy_example}
	\vspace{-15pt}
\end{figure}

The bipartite graph is built upon observed user-item interactions, thus containing the collaborative filtering signal.
Specifically, the first-hop neighborhood directly profiles ego user and item nodes --- \ie historical items of a user (or interacted users of an item) can be viewed as the pre-existing features of user (or item). The second-hop neighboring nodes of a user (or an item) exhibit similar users \wrt behaviors (or similar items \wrt audiences).
Furthermore, the higher-order paths from a user to an item reflect potential interests of the user on the item.
Undoubtedly, mining the inherent patterns in graph structure is helpful to representation learning.
We hence devise three operators on the graph structure, node dropout, edge dropout and random walk, to create different views of nodes.
The operators can be uniformly expressed as follows:
\begin{gather}\label{eq:augmentation_graph}
    \Mat{Z}_1^{(l)}=H(\Mat{Z}_1^{(l-1)},s_1(\Set{G})),\Mat{Z}_2^{(l)}=H(\Mat{Z}_2^{(l-1)},s_2(\Set{G})),s_1,s_2\sim\Set{S},
\end{gather}
where two stochastic selections $s_1$ and $s_2$ are independently applied on graph $\Set{G}$, and establish two correlated views of nodes $\Mat{Z}_1^{(l)}$ and $\Mat{Z}_2^{(l)}$.
We elaborate the augmentation operators as follows:
\begin{itemize}[leftmargin=*]
    \item \textbf{Node Dropout (ND).} With the probability $\rho$, each node is discarded from the graph, together with its connected edges. In particular, $s_1$ and $s_2$ can be modeled as:
    \begin{gather}
        s_1(\Set{G}) = (\Mat{M}'\odot\Set{V},\Set{E}),\quad s_2(\Set{G}) = (\Mat{M}''\odot\Set{V},\Set{E}),
    \end{gather}
    where $\Mat{M}',\Mat{M}''\in\{0,1\}^{|\Set{V}|}$ are two masking vectors which are applied on the node set $\Set{V}$ to generate two subgraphs.
    As such, this augmentation is expected to identify the influential nodes from differently augmented views, and make the representation learning less sensitive to structure changes.

    \item \textbf{Edge Dropout (ED).} It drops out the edges in graph with a dropout ratio $\rho$. Two independent processes are represented as:
    \begin{gather}
        s_1(\Set{G}) = (\Set{V},\Mat{M}_1\odot\Set{E}),\quad s_2(\Set{G}) = (\Set{V},\Mat{M}_2\odot\Set{E}),
    \end{gather}
    where $\Mat{M}_1,\Mat{M}_2\in\{0,1\}^{|\Set{E}|}$ are two masking vectors on the edge set $\Set{E}$. Only partial connections within the neighborhood contribute to the node representations. As such, coupling these two subgraphs together aims to capture the useful patterns of the local structures of a node, and further endows the representations more robustness against noisy interactions.
    
    \item \textbf{Random Walk (RW)}.
    The above two operators generate a subgraph shared across all the graph convolution layers. To explore higher capability, we consider assigning different layers with different subgraphs. This can be seen as constructing an individual subgraph for each node with random walk~\cite{gcc} (see Figure~\ref{fig:toy_example} as an example). 
    Assuming we choose edge dropout at each layer (with different ratio or random seeds), we can formulate RW by making the masking vector to be layer sensitive: 
    \begin{gather}
        s_1(\Set{G}) = (\Set{V},\Mat{M}_1^{(l)}\odot\Set{E}),\quad s_2(\Set{G}) = (\Set{V},\Mat{M}_2^{(l)}\odot\Set{E}),
    \end{gather}
    where $\Mat{M}_1^{(l)},\Mat{M}_2^{(l)}\in\{0,1\}^{|\Set{E}|}$ are two masking vectors on the edge set $\Set{E}$ at layer $l$.
\end{itemize}

We apply these augmentations on graph structure per epoch for simplicity --- that is, we generate two different views of each node at the beginning of a new training epoch (for RW, two different views are generated at each layer).
Note that the dropout and masking ratios remain the same for two independent processes (\ie $s_1$ and $s_2$). We leaving the tuning of different ratios in future work.
It is also worthwhile mentioning that only dropout and masking operations are involved, and no any model parameters are added.

\subsection{Contrastive Learning}
Having established the augmented views of nodes, we treat the views of the same node as the positive pairs (\ie $\{(\Mat{z}'_{u},\Mat{z}''_{u})|u\in\Set{U}\}$), and the views of any different nodes as the negative pairs (\ie $\{(\Mat{z}'_{u},\Mat{z}''_{v})|u,v\in\Set{U},u\neq v\}$).
The auxiliary supervision of positive pairs encourages the consistency  between different views of the same node for prediction, while the supervision of negative pairs enforces the divergence among different nodes.
Formally, we follow SimCLR~\cite{SimCLR} and adopt the contrastive loss, InfoNCE~\cite{InfoNCE}, to maximize the agreement of positive pairs and minimize that of negative pairs:
\begin{gather}\label{eq:self-supervised-loss}
    \Lapl^{user}_{ssl}=\sum_{u\in\Set{U}}-\log\frac{\exp(s(\Mat{z}'_{u},\Mat{z}''_{u})/\tau)}{\sum_{v\in\Set{U}}\exp(s(\Mat{z}'_{u},\Mat{z}''_{v})/\tau)},
\end{gather}
where $s(\cdot)$ measures the similarity between two vectors, which is set as cosine similarity function;
$\tau$ is the hyper-parameter, known as the $temperature$ in softmax.
Analogously, we obtain the contrastive loss of the item side $\Lapl^{item}_{ssl}$.
Combining these two losses, we get the objective function of self-supervised task as $\Lapl_{ssl}=\Lapl^{user}_{ssl}+\Lapl^{item}_{ssl}$.

\subsection{Multi-task Training}
To improve recommendation with the SSL task, we leverage a multi-task training strategy to jointly optimize the classic recommendation task (\cf Equation~\eqref{eq:supervised-loss}) and the self-supervised learning task (\cf Equation~\eqref{eq:self-supervised-loss})
\begin{gather}\label{eq:total_loss}
    \Lapl=\Lapl_{main} + \lambda_{1}\Lapl_{ssl} + \lambda_{2}\norm{\Theta}^{2}_{2},
\end{gather}
where $\Theta$ is the set of model parameters in $L_{main}$ since $L_{ssl}$ introduces no additional parameters; $\lambda_{1}$ and $\lambda_{2}$ are hyperparameters to control the strengths of SSL and $L_{2}$ regularization, respectively.
We also consider the alternative optimization --- pre-training on $\Lapl_{ssl}$ and fine-tuning on $\Lapl_{main}$.
See more details in Section~\ref{sec:pretrain}.

\subsection{Theoretical Analyses of SGL}\label{sec:hard_neg_mining}
In this section, we offer in-depth analyses of SGL, aiming to answer the question: how does the recommender model benefit from the SSL task?
Towards this end, we probe the self-supervised loss in Equation~\eqref{eq:self-supervised-loss} and find one reason:
it has the intrinsic ability to perform hard negative mining, which contributes large and meaningful gradients to the optimization and guides the node representation learning.
In what follows, we present our analyses step by step.

\begin{figure}[t]
 \centering
 \subcaptionbox{$g(x), \tau=1$\label{fig:g_x_1}}{
 \vspace{-5pt}
  \includegraphics[width=0.23\textwidth]{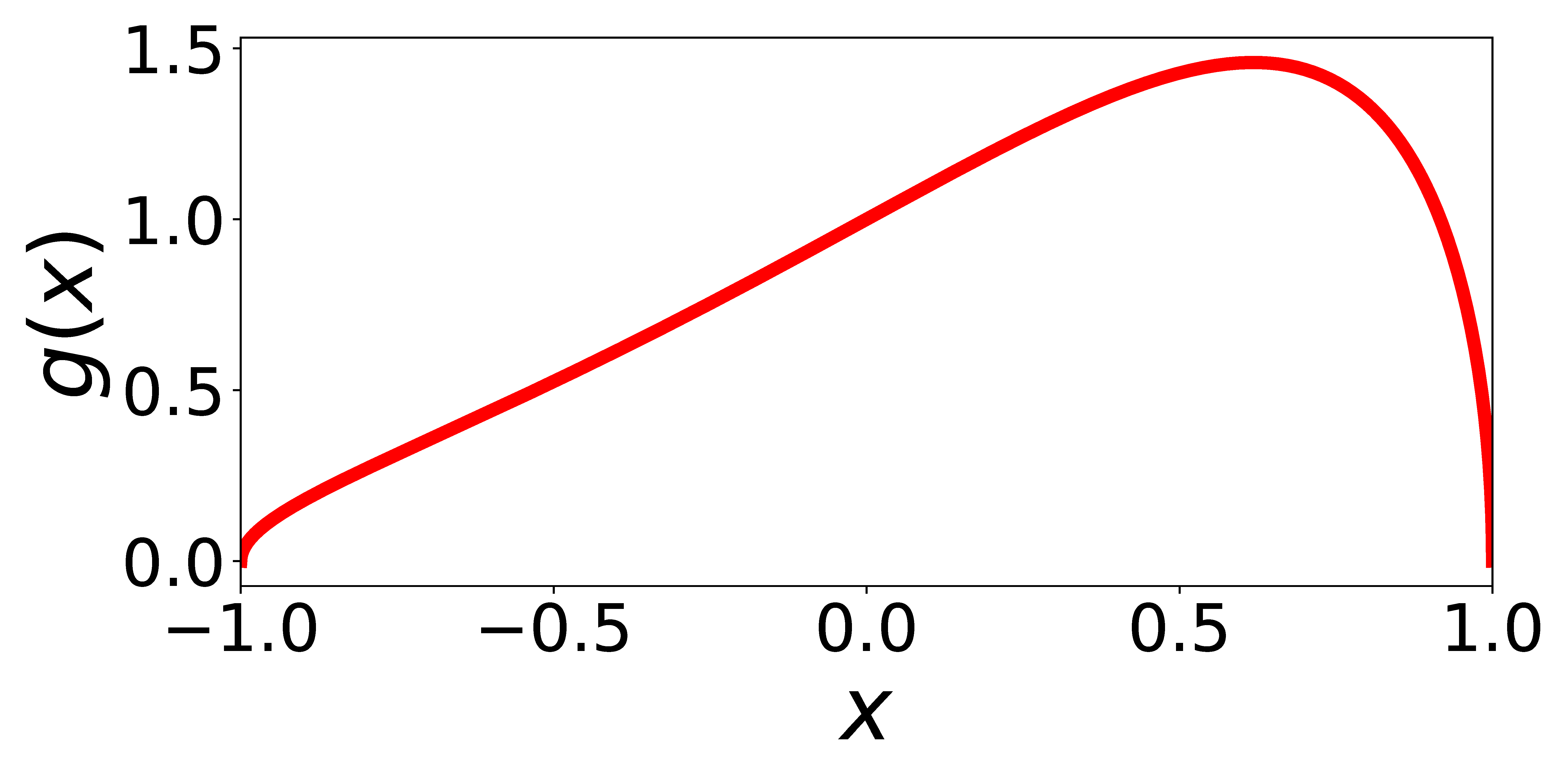}}
  \subcaptionbox{$g(x), \tau=0.1$\label{fig:g_x_2}}{
  \vspace{-5pt}
  \includegraphics[width=0.23\textwidth]{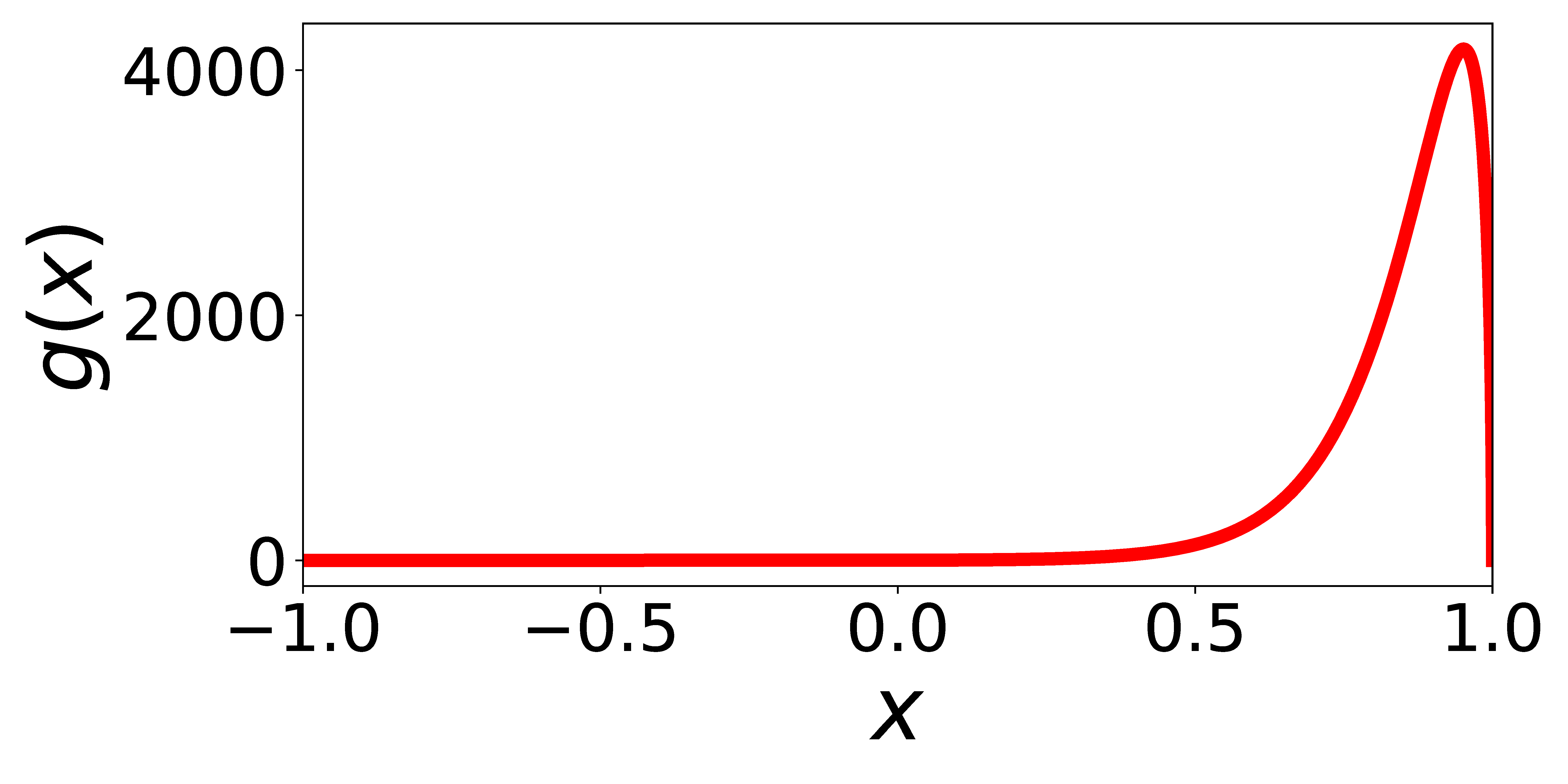}}
 \subcaptionbox{$x^{*}(\tau$)\label{fig:x_tau}}{
 \vspace{-5pt}
  \includegraphics[width=0.23\textwidth]{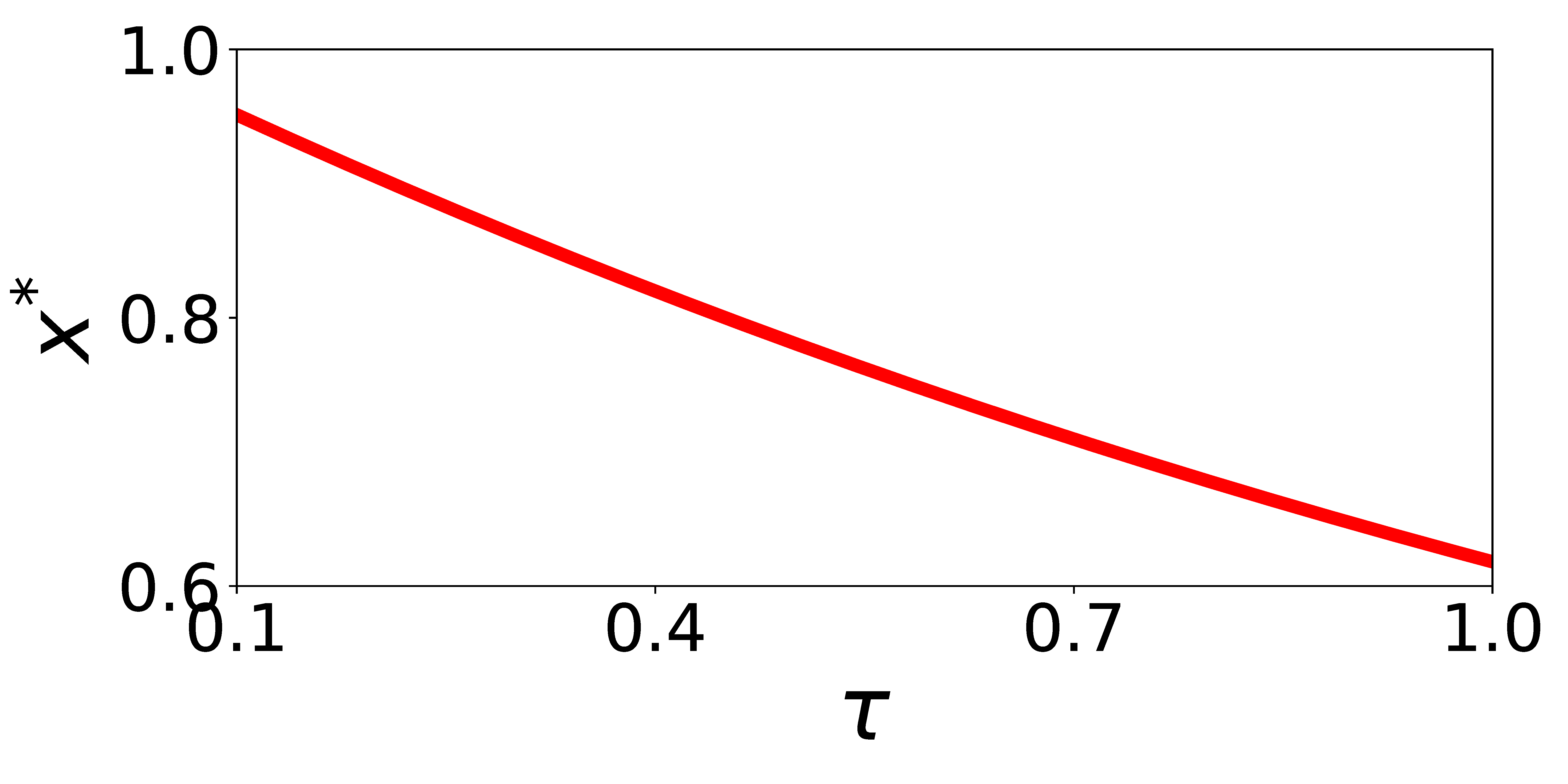}}
  \subcaptionbox{$\ln g(x^{*})$\label{fig:h_tau}}{
  \vspace{-5pt}
  \includegraphics[width=0.23\textwidth]{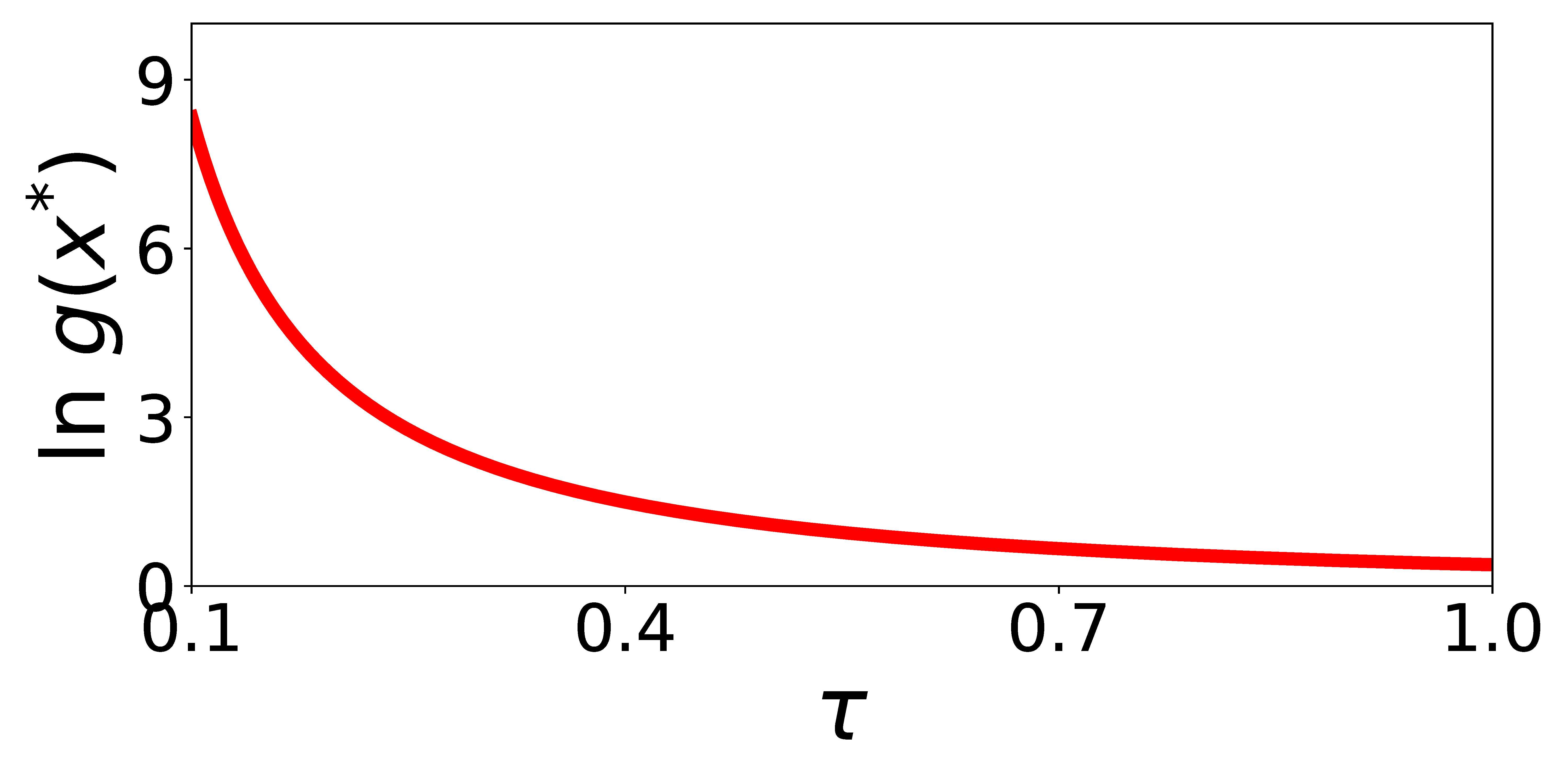}}
  \vspace{-10pt}
  \caption{Function curve of $g(x)$ when $\tau=1$ and $\tau=0.1$, together with the logarithm of the maximum value of $g(x)$ \wrt $\tau$ and its optimal position, \ie $\ln g(x^{*})$ and $x^{*}(\tau)$.}
 \label{fig:function_curve}
 \vspace{-10pt}
\end{figure}

Formally, for node $u\in\Set{U}$, the gradient of the self-supervised loss  \wrt the representation $z_u^{'}$ is as follows:
\begin{align}
    \frac{\partial \Lapl_{ssl}^{user}(u)}{\partial z'_u}
    = \frac{1}{\tau \left \| z'_u \right \|}\Big\{c(u) + \sum_{v\in\Set{U}\setminus\{u\}} c(v)\Big\},
    \label{Eq:gradients}
\end{align}
where
$\Lapl_{ssl}^{user}(u)$ is the individual term for a single node $u$ in Equation~\eqref{eq:self-supervised-loss};
$v\in\Set{U}\setminus \{u\}$ is another node which serves as the negative view for node $u$; $c(u)$ and $c(v)$ separately represent the contribution of positive node $u$ and negative nodes $\{v\}$ to the gradients \wrt $z'_u$:
\begin{align}
    &c(u)
    =  \left( s''_u - ({s'_u}^Ts''_u)s'_u \right)^T (P_{uu} - 1),\\
    &c(v)
    = \left( s''_v - ({s'_u}^Ts''_v)s'_u \right)^T P_{uv},\label{equ:negative-contribution}
\end{align}
where $P_{uv} = \frac{\exp({s_u^{'}}^T s_v^{''}/\tau)}{\sum\limits_{v\in \Set{U}} \exp({s'_u}^T s''_v / \tau)}$; $s'_{u} = \frac{z'_{u}}{\left \| z'_{u} \right \|}$ and $ s''_{u} = \frac{z''_{u}}{\left \| z''_{u} \right \|}$ are the normalized representations of node $u$ from different views; similar notations to node $v$.
Afterwards, we focus on the contribution of negative node $v$ (\cf Equation \eqref{equ:negative-contribution}), the $L_{2}$ norm of which is proportional to the following term:
\begin{gather}\label{eq:negative-contribution-prop}
    {\left \| c(v) \right\|}_2 \propto\sqrt{1 - ({s_u^{'}}^Ts_v^{''})^2} \exp ({s_u^{'}}^T s_v^{''} / \tau).
\end{gather}
As $s_u^{'}$ and $s_v^{''}$ are both unit vectors, we can introduce another variable $x = {s_u^{'}}^Ts_v^{''} \in [-1,1]$ to simplify Equation \eqref{eq:negative-contribution-prop} as follows:
\begin{gather}
    g(x)=\sqrt{1-x^2} \exp \left( \frac{x}{\tau} \right),
\end{gather}
where $x$ directly reflects the representation similarity between the positive node $u$ and the negative node $v$.
According to the similarity $x$, we can roughly categorize the negative nodes into two groups:
(1) Hard negative nodes, whose representations are similar to that of the positive node $u$ (\ie $0 < x \leq 1$), thus making it difficult to distinguish $v$ from $u$ in the latent space; (2) Easy negative nodes, which are dissimilar to the positive node $u$ (\ie $-1 \leq x < 0$) and can be easily discriminated.
To investigate the contributions of hard and easy negatives, we plot the curves of $g(x)$ over the change of the node similarity $x$ in Figures \ref{fig:g_x_1} and \ref{fig:g_x_2}, by setting $\tau=1$ and $\tau=0.1$ respectively.
Clearly, under different conditions of $\tau$, the contributions of negative nodes differ dramatically with each others.
Specifically, as Figure \ref{fig:g_x_1} shows, given $\tau=1$, the values of $g(x)$ fall into the range of $(0,1.5)$ and slightly change in response to $x$.
This suggests that negative samples, no matter hard or easy, contribute similarly to the gradient.
In contrast, as Figure \ref{fig:g_x_2} displays, when setting $\tau=0.1$, the values of $g(x)$ at hard negatives could reach $4,000$, while the contribution of easy negatives is vanishing.
This indicates that hard negative nodes offer much larger gradients to guide the optimization, thus making node representations more discriminative and accelerating the training process \cite{DBLP:conf/wsdm/RendleF14}.

These findings inspire us to probe the influence of $\tau$ on the maximum value of $g(x)$.
By approaching $g(x)$'s derivative to zero, we can obtain $x^{*}={(\sqrt{\tau^2 + 4} - \tau)}/{2}$ with maximum value $g(x^{*})$.
To see how $g(x^{*})$ changes with $\tau$,
we represent its logarithm as:
\begin{align}
    \ln g(x^{*}) = \ln \left( \sqrt{ 1 - \left( \frac{\sqrt{\tau^2 + 4}-\tau}{2} \right)^2} \exp{\left( \frac{\sqrt{\tau^2 + 4}  - \tau}{2\tau} \right)} \right).
\end{align}
We present the curves of $x^{*}$ and $\ln g(x^{*})$ in Figures \ref{fig:x_tau} and \ref{fig:h_tau}, respectively.
With the decrease of $\tau$, the most influential negative nodes become more similar to the positive node (\ie $x^{*}$ approaches $0.9$), moreover, their contributions are amplified super-exponentially (\ie $g^{*}$ is close to $e^{8}$).
Hence, properly setting $\tau$ enables SGL to automatically perform hard negative mining.

It is worth mentioning that, 
our analyses are inspired by the prior study \cite{KhoslaTWSTIMLK20}, but there are major differences:
(1) \cite{KhoslaTWSTIMLK20} defines hard negatives as samples whose similarity are near-to-zero to the positive sample (\ie $x\approx 0$), and easy negatives as obviously dissimilar samples (\ie $x\approx -1$);
(2) It leaves the region of $x>0$ untouched, which is however crucial in our case.
Innovatively, we provide a more fine-grained view in the region of $x>0$ to highlight the critical role of $\tau$, the temperature hyper-parameter in softmax, in mining hard negatives.

\subsection{Complexity Analyses of SGL}
In this subsection, we analyze the complexity of SGL with ED as the strategy and LightGCN as the recommendation model; other choices can be analyzed similarly. 
Since SGL introduces no trainable parameters, the space complexity remains the same as LightGCN~\cite{LightGCN}.
The time complexity of model inference is also the same, since there is no change on the model structure. 
In the following part, we will analyze the time complexity of SGL training.

Suppose the number of nodes and edges in the user-item interaction graph are $\left| V \right|$ and $\left| E \right|$ respectively. Let $s$ denote the number of epochs, $B$ denote the size of each training batch, $d$ denote the embedding size, $L$ denote the number of GCN layers, $\hat{\rho}=1-\rho$ denote the keep probability of SGL-ED. The complexity mainly comes from two parts:
\begin{itemize}[leftmargin=*]
    \item Normalization of adjacency matrix. Since we generate two independent sub-graphs per epoch, given the fact that the number of non-zero elements in the adjacency matrices of full training graph and two sub-graph are $2 \left| E \right|, 2 \hat{\rho} \left| E \right|$ and $2 \hat{\rho} \left| E \right|$ respectively, its total complexity is 
    $O(4 \hat{\rho} \left| E \right| s + 2 \left| E \right|)$.
    
    \item Evaluating self-supervised loss. We only consider the inner product in our analyses. As defined in Equation~(\ref{eq:self-supervised-loss}), we treat all other user nodes as negative samples when calculating InfoNCE loss of user side. Within a batch, the complexity of numerator and denominator are $O(B d)$ and $O(B M d)$, respectively, where $M$ is the number of users. And hence the total complexity of both user and item side per epoch is $O(\left| E \right| d (2 + \left| V \right|))$. Therefore, the time complexity of the whole training phase is $O(\left| E \right| d (2 + \left| V \right|) s)$. An alternative to reduce the time complexity is treating only the users (or the items) within the batch as negative samples~\cite{SimCLR,DBLP:journals/corr/abs-2007-12865}, resulting in total time complexity of $O(\left| E \right| d (2 + 2B) s)$.
\end{itemize}
We summarize the time complexity in training between LightGCN and SGL-ED in Table~\ref{tab:time-complexity}.
The analytical complexity of LightGCN and SGL-ED is actually in the same magnitude, since the increase of LightGCN only scales the complexity of LightGCN.
In practice, taking the Yelp2018 data as an example, the time complexity of SGL-ED (alternative) with $\hat{\rho}$ of 0.8 is about 3.7x larger than LightGCN, which is totally acceptable considering the speedup of convergence speed we will show in Section~\ref{sec:exp_efficiency}.
The testing platform is Nvidia Titan RTX graphics card equipped with Inter i7-9700K CPU (32GB Memory). The time cost of each epoch on Yelp2018 is 15.2s and 60.6s for LightGCN and SGL-ED (alternative) respectively, which is consistent with the complexity analyses.

\begin{table}[t]
\caption{The comparison of analytical time complexity between LightGCN and SGL-ED.}
\vspace{-10px}
\label{tab:time-complexity}
\resizebox{0.46\textwidth}{!}{
\begin{tabular}{c|c|c}
\hline
Component                                             & LightGCN & SGL-ED \\ \hline\hline
\begin{tabular}[c]{@{}c@{}}Adjacency\\ Matrix\end{tabular}     
    & $O(2 \left| E \right|)$                  
    & $O(4 \hat{\rho} \left| E \right| s + 2 \left| E \right|)$                \\ \hline
\begin{tabular}[c]{@{}c@{}}Graph\\ Convolution\end{tabular}    
    & $O(2 \left| E \right| L d s \frac{\left| E \right|}{B})$                  
    & $O(2(1+2\hat{\rho}) \left| E \right| L d s \frac{\left| E \right|}{B})$                \\ \hline
\begin{tabular}[c]{@{}c@{}}BPR\\ Loss\end{tabular}             
    & $O(2 \left| E \right| d s)$                  
    & $O(2 \left| E \right| d s)$                \\ \hline
\multirow{2}{*}{\begin{tabular}[c]{@{}c@{}}Self-supervised\\ Loss\end{tabular}} & \multirow{2}{*}{-} 
& $O(\left| E \right| d (2 + \left| V \right|) s)$           \\ \cline{3-3} 
&                    
& $O(\left| E \right| d (2 + 2B) s)$           \\ \hline
\end{tabular}}
\vspace{-5px}
\end{table}
\section{Experiments}

\begin{table}
\centering
\caption{Statistics of the datasets.}
\vspace{-10pt}
\label{Table:Statistics of the experimented data}
\resizebox{0.46\textwidth}{!}{
\begin{tabular}{l|r|r|r|r}
\hline
Dataset          & \#Users & \#Items & \#Interactions & Density \\ \hline\hline
Yelp2018         & 31,668  & 38,048  & 1,561,406      & 0.00130 \\ 
Amazon-Book      & 52,643  & 91,599  & 2,984,108      & 0.00062 \\ 
Alibaba-iFashion & 300,000 & 81,614  & 1,607,813      & 0.00007 \\ \hline
\end{tabular}}
\vspace{-10pt}
\end{table}

To justify the superiority of SGL and reveal the reasons of its effectiveness, we conduct extensive experiments
and answer the following research questions:
\begin{itemize}[leftmargin=*]
    \item \textbf{RQ1:} How does SGL perform \wrt top-$K$ recommendation as compared with the state-of-the-art CF models?
    \item \textbf{RQ2:} What are the benefits of performing self-supervised learning in collaborative filtering?
    \item \textbf{RQ3:} How do different settings influence the effectiveness of the proposed SGL?
\end{itemize}

\begin{table*}[t]
\centering
\caption{Performance comparison with LightGCN at different layers. The performance of LightGCN on Yelp2018 and Amazon-Book are copied from its original paper. The percentage in brackets denote the relative performance improvement over LightGCN. The bold indicates the best result.}
\vspace{-10pt}
\label{Table:comparion_with_LightGCN}
\resizebox{0.95\textwidth}{!}{
\begin{tabular}{l|l|l|l|l|l|ll}
\hline
\multicolumn{2}{c|}{\textbf{Datase}}                                         & \multicolumn{2}{c|}{\textbf{Yelp2018}}                                    & \multicolumn{2}{c|}{\textbf{Amazon-Book}}                                    & \multicolumn{2}{c}{\textbf{Alibaba-iFashion}}                                     \\ \hline
\multicolumn{1}{c|}{\textbf{\#Layer}} & \multicolumn{1}{c|}{\textbf{Method}} & \multicolumn{1}{c|}{\textbf{Recall}} & \multicolumn{1}{c|}{\textbf{NDCG}} & \multicolumn{1}{c|}{\textbf{Recall}} & \multicolumn{1}{c|}{\textbf{NDCG}} & \multicolumn{1}{c|}{\textbf{Recall}}          & \multicolumn{1}{c}{\textbf{NDCG}} \\ \hline
\multirow{4}{*}{\textbf{1 Layer}}     & LightGCN                             & 0.0631                               & 0.0515                             & 0.0384                               & 0.0298                             & \multicolumn{1}{l|}{0.0990}                   & 0.0454                            \\
                                      & SGL-ND                               & \textbf{0.0643(+1.9\%)}              & \textbf{0.0529(+2.7\%)}            & 0.0432(+12.5\%)                      & 0.0334(+12.1\%)                    & \multicolumn{1}{l|}{\textbf{0.1133(+14.4\%)}} & \textbf{0.0539(+18.7\%)}          \\
                                      & SGL-ED                               & 0.0637(+1.0\%)                       & 0.0526(+2.1\%)                     & \textbf{0.0451(+17.4\%)}             & \textbf{0.0353(+18.5\%)}           & \multicolumn{1}{l|}{0.1125(+13.6\%)}          & 0.0536(+18.1\%)                   \\
                                      & SGL-RW                               & 0.0637(+1.0\%)                       & 0.0526(+2.1\%)                     & \textbf{0.0451(+17.4\%)}             & \textbf{0.0353(+18.5\%)}           & \multicolumn{1}{l|}{0.1125(+13.6\%)}          & 0.0536(+18.1\%)                   \\ \hline
\multirow{4}{*}{\textbf{2 Layers}}    & LightGCN                             & 0.0622                               & 0.0504                             & 0.0411                               & 0.0315                             & \multicolumn{1}{l|}{0.1066}                   & 0.0505                            \\
                                      & SGL-ND                               & 0.0658(+5.8\%)                       & 0.0538(+6.7\%)                     & 0.0427(+3.9\%)                       & 0.0335(+6.3\%)                     & \multicolumn{1}{l|}{\textbf{0.1106(+3.8\%)}}  & \textbf{0.0526(+4.2\%)}           \\
                                      & SGL-ED                               & \textbf{0.0668(+7.4\%)}              & \textbf{0.0549(+8.9\%)}            & \textbf{0.0468(+13.9\%)}             & \textbf{0.0371(+17.8\%)}           & \multicolumn{1}{l|}{0.1091(+2.3\%)}           & 0.0520(+3.0\%)                    \\
                                      & SGL-RW                               & 0.0644(+3.5\%)                       & 0.0530(+5.2\%)                     & 0.0453(+10.2\%)                      & 0.0358(+13.7\%)                    & \multicolumn{1}{l|}{0.1091(+2.3\%)}           & 0.0521(+3.2\%)                    \\ \hline
\multirow{4}{*}{\textbf{3 Layers}}    & LightGCN                             & 0.0639                               & 0.0525                             & 0.0410                               & 0.0318                             & \multicolumn{1}{l|}{0.1078}                   & 0.0507                            \\
                                      & SGL-ND                               & 0.0644(+0.8\%)                       & 0.0528(+0.6\%)                     & 0.0440(+7.3\%)                       & 0.0346(+8.8\%)                     & \multicolumn{1}{l|}{0.1126(4.5\%)}            & 0.0536(+5.7\%)                    \\
                                      & SGL-ED                               & \textbf{0.0675(+5.6\%)}              & \textbf{0.0555(+5.7\%)}            & \textbf{0.0478(+16.6\%)}             & \textbf{0.0379(+19.2\%)}           & \multicolumn{1}{l|}{0.1126(+4.5\%)}           & 0.0538(+6.1\%)                    \\
                                      & SGL-RW                               & 0.0667(+4.4\%)                       & 0.0547(+4.2\%)                     & 0.0457(+11.5\%)                      & 0.0356(+12.0\%)                    & \multicolumn{1}{l|}{\textbf{0.1139(+5.7\%)}}  & \textbf{0.0539(+6.3\%)}           \\ \hline
\end{tabular}}
\vspace{-10pt}
\end{table*}

\subsection{Experimental Settings}\label{sec:exp_settings}
We conduct experiments on three benchmark datasets: Yelp2018\cite{NGCF,LightGCN}, Amazon-Book\cite{NGCF,LightGCN}, and Alibaba-iFashion~\cite{POG}\footnote{https://github.com/wenyuer/POG}. Following~\cite{NGCF,LightGCN}, we use the same 10-core setting for Yelp2018 and Amazon-Book. Alibaba-iFashion is more sparse, where we randomly sample 300k users and use all their interactions over the fashion outfits. The statistics of all three datasets are summarized in Table~\ref{Table:Statistics of the experimented data}.

We follow the same strategy described in~\cite{NGCF} to split the interactions into training, validation, and testing set with a ratio of 7:1:2. 
For users in the testing set, we follow the all-ranking protocol~\cite{NGCF} to evaluate the top-$K$ recommendation performance and report the average Recall@$K$ and NDCG@$K$ where $K=20$.

\subsubsection{\textbf{Compared Methods}}
We compare the proposed SGL with the following CF models:
\begin{itemize}[leftmargin=*]
    \item NGCF~\cite{NGCF}. This is a graph-based CF method largely follows the standard GCN~\cite{GCN}.
    It additionally encodes the second-order feature interaction into the message during message passing. We tune the regularization coefficient $\lambda_{2}$ and the number of GCN layers within the suggested ranges. 
    \item LightGCN~\cite{LightGCN}.
    This method devises a light graph convolution for training efficiency and generation ability. Similarly, we tune the $\lambda_{2}$ and the number of GCN layers.
    \item Mult-VAE~\cite{MultiVAE}. This is an item-based CF method based on the variational auto-encoder (VAE).
    It is optimized with an additional reconstruction objective, which can be seen as a special case of SSL. We follow the suggested model setting and tune the dropout ratio and $\beta$.
    \item DNN+SSL~\cite{DBLP:journals/corr/abs-2007-12865}. This is a state-of-the-art SSL-based recommendation method. With DNNs as the encoder of items, it adopts two augmentation operators, feature masking (FM) and feature dropout (FD), on the pre-existing features of items. In our cases where no item feature is available, we apply the augmentations  on ID embeddings of items instead. We tune the DNN architecture (\ie the number of layers and the number of neurons per layer) as suggested in the original paper.
\end{itemize}

We discard potential baselines like MF~\cite{BPR}, NeuMF~\cite{NeuMF}, GC-MC~\cite{GCMC}, and PinSage~\cite{PinSage} since the previous work~\cite{NGCF,LightGCN,MultiVAE} has validated the superiority over the compared ones. Upon LightGCN, we implement three variants of the proposed SGL, named SGL-ND, SGL-ED, and SGL-RW, which are equipped with Node Dropout, Edge Dropout, and Random Walk, respectively.

\subsubsection{\textbf{Hyper-parameter Settings}}\label{sec:parameter_settings}
For fair comparison, all models are trained from scratch which are initialized with the Xavier method~\cite{Xavier}. The models are optimized by the Adam
optimizer with learning rate of 0.001 and mini-batch size of 2048. The early stopping strategy is the same as NGCF and LightGCN. 
The proposed SGL methods inherit the optimal values of the shared hyper-parameters.
For the unique ones of SGL, we tune $\lambda_{1}$, $\tau$, and $\rho$ within the ranges of $\{0.005,0.01,0.05,0.1,0.5,1.0\}$, $\{0.1,0.2,0.5,1.0\}$, and $\{0,0.1,0.2,\cdots,0.5\}$, respectively. 

\begin{figure*}[th]
 \centering
 \subcaptionbox{Yelp2018\label{fig:long-tail-yelp}}{
  \includegraphics[width=0.3\textwidth]{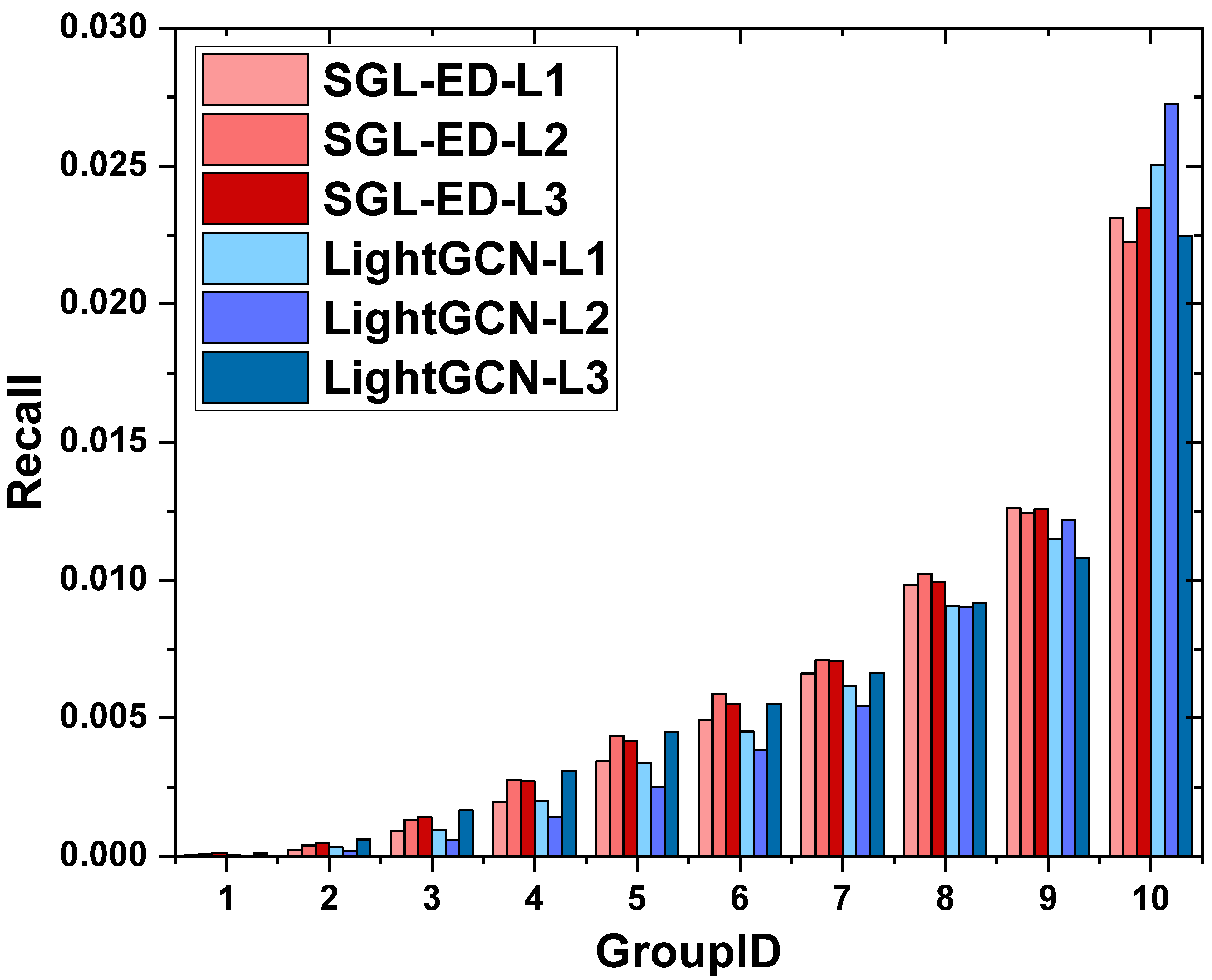}}
 \subcaptionbox{Amazon-Book\label{fig:long-tail-ab}}{
  \includegraphics[width=0.3\textwidth]{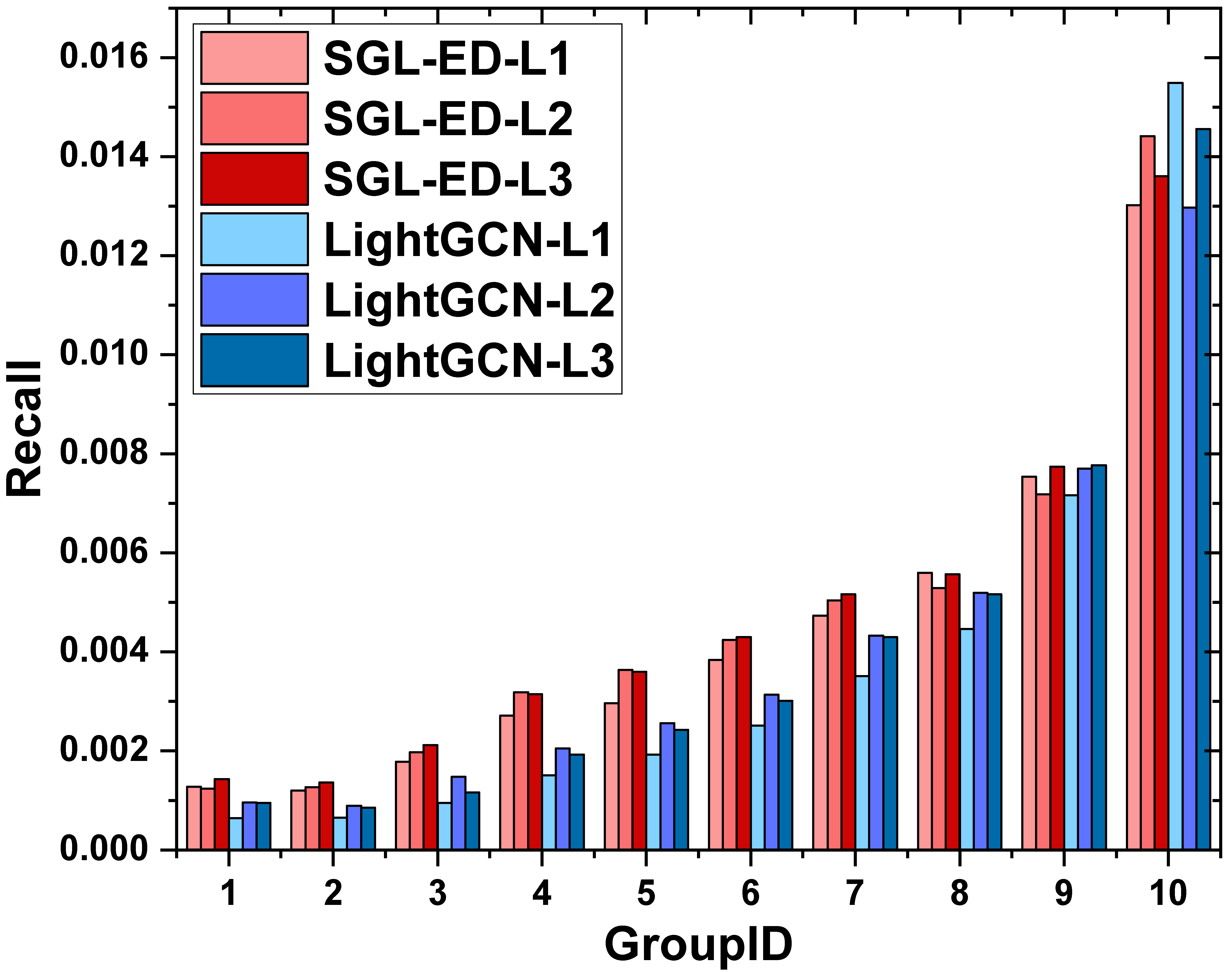}}
  \subcaptionbox{Alibaba-iFashion\label{fig:long-tail-ifashion}}{
  \includegraphics[width=0.3\textwidth]{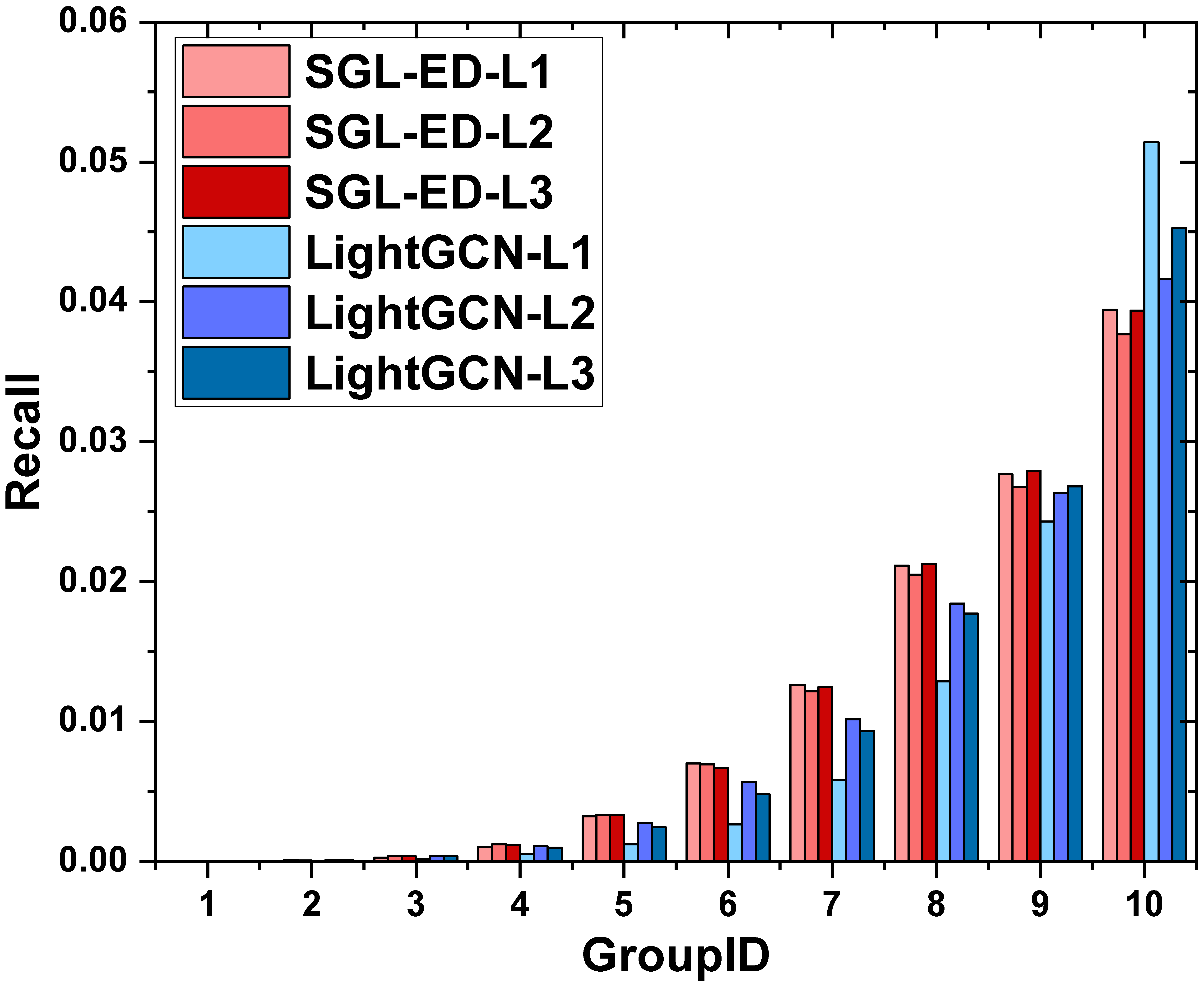}}
  \vspace{-10pt}
  \caption{Performance comparison over different item groups between SGL-ED and LightGCN. The suffix in the legend indicates the number of GCN layers.}
 \label{fig:long-tail}
 \vspace{-10pt}
\end{figure*}

\subsection{Performance Comparison (RQ1)}\label{sec:overall_performance}

\subsubsection{\textbf{Comparison with LightGCN}} 
Table~\ref{Table:comparion_with_LightGCN} shows the result comparison between SGL and LightGCN.
We find that:
\begin{itemize}[leftmargin=*]
    \item In most cases, three SGL implementations outperform LightGCN by a large margin, indicating the superiority of supplementing the recommendation task with self-supervised learning.
    
    \item In SGL family, SGL-ED achieves best performance in 10 over 18 cases, while SGL-RW also performs better than SGL-ND across all three datasets.
    We ascribe these to the ability of edge dropout-like operators to capture inherent patterns in graph structure.
    Moreover, the performance of SGL-ED is better than that of SGL-RW in the denser datasets (Yelp2018 and Amazon-Book), while slightly worse in the sparser dataset (Alibaba-iFashion).
    One possible reason is that, in sparser datasets, ED is more likely to blocks connections of low-degree nodes (inactive users and unpopular items), while RW could restore their connections at different layers, as the case of node $u_1$ and $i_1$ shown in Figure~\ref{fig:toy_example}.
    
    \item SGL-ND is relatively unstable than SGL-ED and SGL-RW. For example, on Yelp2018 and Amazon-Book, the results of SGL-ED and SGL-RW increases when layers go deeper while SGL-ND exhibits different patterns.
    Node Dropout can be viewed as a special case of Edge Dropout, which discards edges around a few nodes.
    Hence, dropping high-degree nodes will dramatically change the graph structure, thus exerts influence on the information aggregation and makes the training unstable.
    
    \item The improvements on Amazon-Book and Alibaba-iFashion are more significant than that on Yelp2018. This might be caused by the characteristics of datasets.
    Specifically, in Amazon-Book and Alibaba-iFashion, supervision signal from user-item interactions is too sparse to guide the representation learning in LightGCN.
    Benefiting from the self-supervised task, SGL obtains auxiliary supervisions to assist the representation learning.
    
    \item Increasing the model depth from 1 to 3 is able to enhance the performance of SGL. This indicates that exploiting SSL could empower the generalization ability of GCN-based recommender models --- that is, the contrastive learning among different nodes is of promise to solving the oversmoothing issue of node representations, further avoiding the overfitting problem.
\end{itemize}

\subsubsection{\textbf{Comparison with the State-of-the-Arts}}
In Table~\ref{tab:overall}, we summarize the performance comparison with various baselines.
We find that:
(1) SGL-ED consistently outperforms all baselines across the board.
This again verifies the rationality and effectiveness of incorporating the self-supervised learning.
(2) LightGCN achieves better performance than NGCF and Mult-VAE, which is consistent to the claim of LightGCN paper. The performance of Mult-VAE is on par with NGCF and LightGCN on Alibaba-iFashion, while outperforms NGCF on Amazon-Book.
(3) DNN+SSL is the strongest baseline on Amazon-Book, showing the great potential of SSL in recommendation.
Surprisingly, on the other datasets, DNN+SSL performs much worse than SGL-ED.
This suggests that directly applying SSL on ID embeddings might be suboptimal and inferior than that on graph representations.
(4) Moreover, we conduct the significant test, where $p$-value $< 0.05$ indicates that the improvements of SGL-ED over the strongest baseline are
statistically significant in all six cases.

\begin{table}[t]
\caption{Overall Performance Comparison.}
\label{tab:overall}
\vspace{-10pt}
\resizebox{0.46\textwidth}{!}{
\begin{tabular}{c|cc|cc|cc}
\hline
\textbf{Dataset} & \multicolumn{2}{c|}{Yelp2018} & \multicolumn{2}{c|}{Amazon-Book} & \multicolumn{2}{c}{Alibaba-iFashion} \\ \hline
Method  & Recall    & NDCG     & Recall     & NDCG       & Recall       & NDCG         \\ \hline\hline
NGCF             & 0.0579             & 0.0477            & 0.0344              & 0.0263              & 0.1043                & 0.0486                \\
LightGCN         & \uline{0.0639}             & \uline{0.0525}            & 0.0411              & 0.0315              & \uline{0.1078}                & \uline{0.0507}                \\
Mult-VAE         & 0.0584             & 0.0450            & 0.0407              & 0.0315              & 0.1041                & 0.0497                \\
DNN+SSL          & 0.0483     & 0.0382      & \uline{0.0438}      & \uline{0.0337}       & 0.0712      & 0.0325 \\  \hline
SGL-ED           & \textbf{0.0675}    & \textbf{0.0555}   & \textbf{0.0478}     & \textbf{0.0379}     & \textbf{0.1126}       & \textbf{0.0538}       \\ \hline \hline
\%Improv.       & 5.63\%      & 5.71\%      & 9.13\%      & 12.46\%      & 4.45\%      & 6.11\% \\
$p$-value       & 5.92e-8      & 1.89e-8      & 5.07e-10      & 
3.63e-10      & 3.34e-8      & 4.68e-10 \\ \hline
\end{tabular}}
\vspace{-10pt}
\end{table}

\subsection{Benefits of SGL (RQ2)}~\label{sec:exp_effective}
In this section, we study the benefits of SGL from three dimensions:
(1) long-tail recommendation;
(2) training efficiency;
and (3) robustness to noises.
Due to the limited space, we only report the results of SGL-ED, while having similar observations in others.

\subsubsection{\textbf{Long-tail Recommendation}}
As Introduction mentions, GNN-based recommender models easily suffer from the long-tail problem.
To verify whether SGL is of promise to solving the problem, we split items into ten groups based on the popularity, meanwhile keeping the total number of interactions of each group the same.
The larger the GroupID is, the larger degrees the items have.
We then decompose the Recall@20 metric of the whole dataset into contributions of single groups, as follows:
\begin{align}
    Recall 
    = \frac{1}{M}\sum_{u=1}^{M}\frac{\sum\limits_{g=1}^{10} \left | (l_{rec}^u)^{(g)} \cap l_{test}^u \right |}{\left | l_{test}^u \right |} \nonumber= \sum\limits_{g=1}^{10} Recall^{(g)}
\end{align}
where $M$ is the number of users, $l_{rec}^{u}$ and $l_{test}^{u}$ are the items in the top-$K$ recommendation list and relevant items in the testing set for user $u$, respectively.
As such, $Recall^{(g)}$ measures the recommendation performance over the $g$-th group.
We report the results in Figure~\ref{fig:long-tail} and find that:
\begin{itemize}[leftmargin=*]
    \item LightGCN is inclined to recommend high-degree items, while leaving long-tail items less exposed. Specifically, although only containing 0.83\%, 0.83\% and 0.22\% of item spaces, the 10-th group contributes 39.72\%, 39.92\% and 51.92\% of the total Recall scores in three datasets, respectively.
    This admits that, LightGCN hardly learns high-quality representations of long-tail items, due to the sparse interaction signal.
    Our SGL shows potentials in alleviating this issue: the contributions of the 10-group downgrade to 36.27\%, 29.15\% and 35.07\% in three datasets, respectively.

    \item Jointly analyzing Table.~\ref{Table:comparion_with_LightGCN} and Figure~\ref{fig:long-tail}, we find the performance improvements of SGL mainly come from accurately recommending the items with sparse interactions. This again verifies that the representation learning benefits greatly from auxiliary supervisions, so as to establish better representations of these items than LightGCN.
\end{itemize}

\subsubsection{\textbf{Training Efficiency}}~\label{sec:exp_efficiency}
\begin{figure}[t]
	\centering
	\subcaptionbox{Yelp2018-L3\label{fig:curve-yelp}}{
        \includegraphics[width=0.21\textwidth]{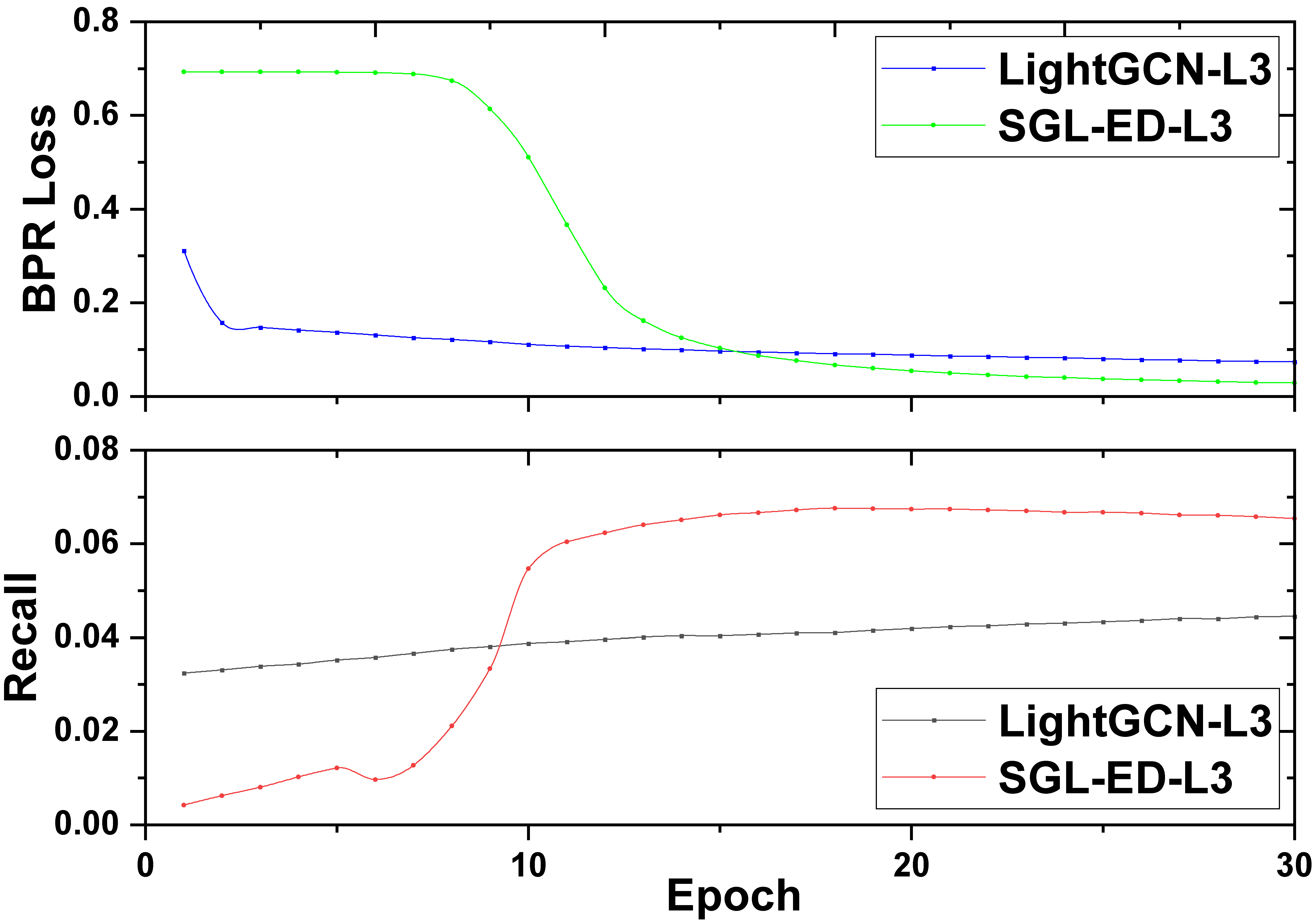}}
    \subcaptionbox{Amazon-Book-L3\label{fig:curve-ab}}{
        \includegraphics[width=0.21\textwidth]{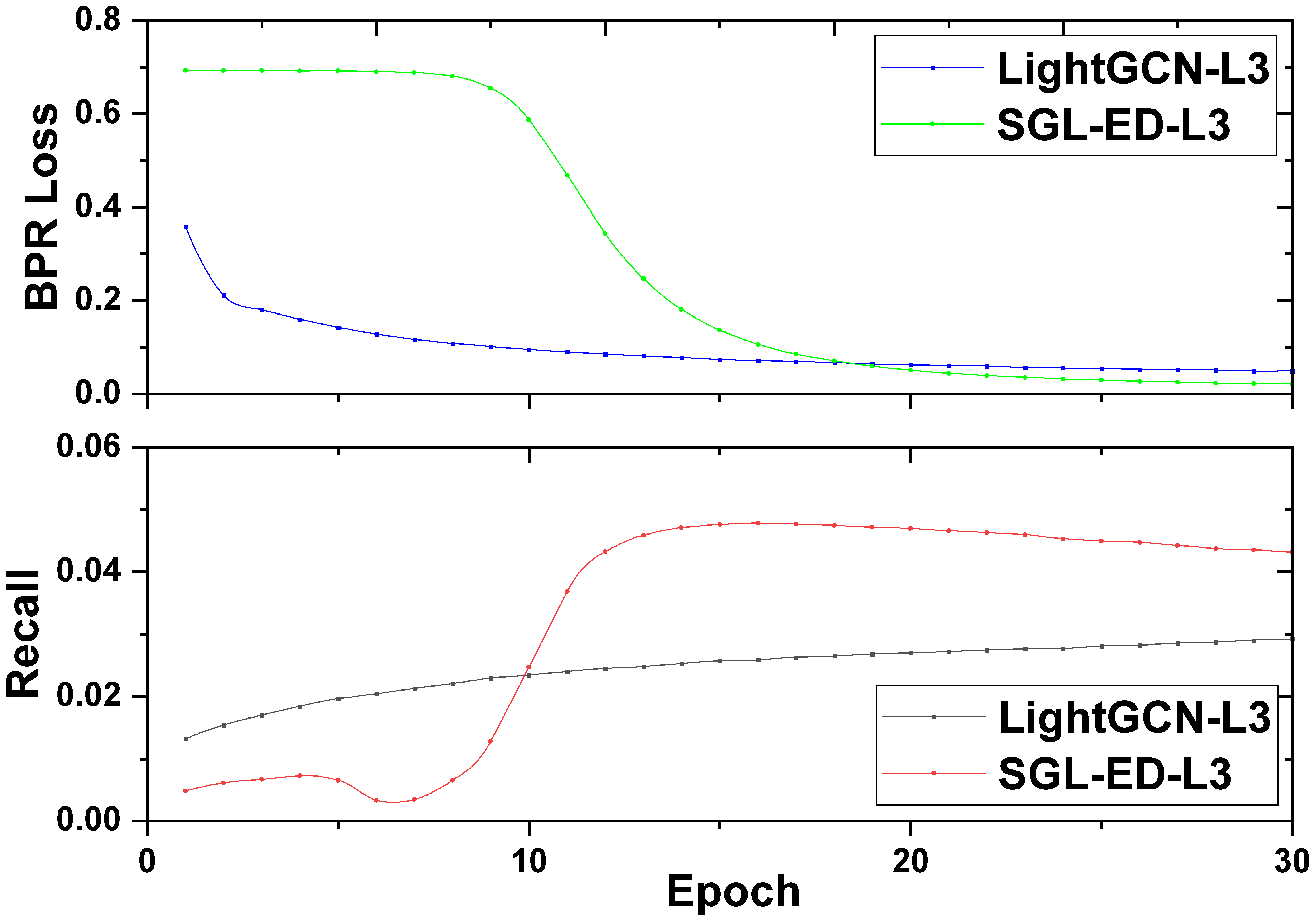}}
    \vspace{-10pt}
	\caption{Training Curves of SGL-ED and LightGCN on three datasets. The suffix in the legend denotes the layer numbers.}
	\label{fig:training_curve}
	\vspace{-10pt}
\end{figure}

Self-supervised learning has proved its superiority in pre-training natural language model~\cite{DBLP:conf/naacl/DevlinCLT19} and graph structure~\cite{gcc,DBLP:conf/iclr/HuLGZLPL20}.
Thus, we would like to study its influence on training efficiency.
Figure~\ref{fig:training_curve} shows the training curves of SGL-ED and LightGCN on Yelp2018 and Amazon-Book. 
As the number of epochs increases, the upper subfigures display the changes of training loss, while the bottoms record the performance changes in the testing set.
We have the following observations:
\begin{itemize}[leftmargin=*]
    \item Obviously, SGL is much faster to converge than LightGCN on Yelp2018 and Amazon-Book. In particular, SGL arrives at the best performance at the 18-th and 16-th epochs, while LightGCN takes 720 and 700 epochs in these two datasets respectively. This suggests that our SGL can greatly reduce the training time, meanwhile achieves remarkable improvements.
    We ascribe such speedups to two factors: (1) SGL adopts the InfoNCE loss as the SSL objective which enables the model to learn representations from multiple negative samples, while the BPR loss in LightGCN only uses one negative sample which limits the model's perception field; (2) As analyzed in Section~\ref{sec:hard_neg_mining}, with a proper $\tau$, SGL benefits from dynamic hard negative mining, where the hard negative samples offer meaningful and larger gradients to guide the optimization~\cite{DNS}.
    
    \item Another observation is that the origin of the rapid-decline period of BPR loss is slightly later than the rapid-rising period of Recall. This indicates the existence of a gap between the BPR loss and ranking task.
    We will conduct in-depth research on this phenomenon in our future work.
\end{itemize}

\subsubsection{\textbf{Robustness to Noisy Interactions}}

\begin{figure}[t]
	\centering
	\subcaptionbox{Yelp2018\label{fig:denoise_yelp}}{
		\includegraphics[width=0.23\textwidth]{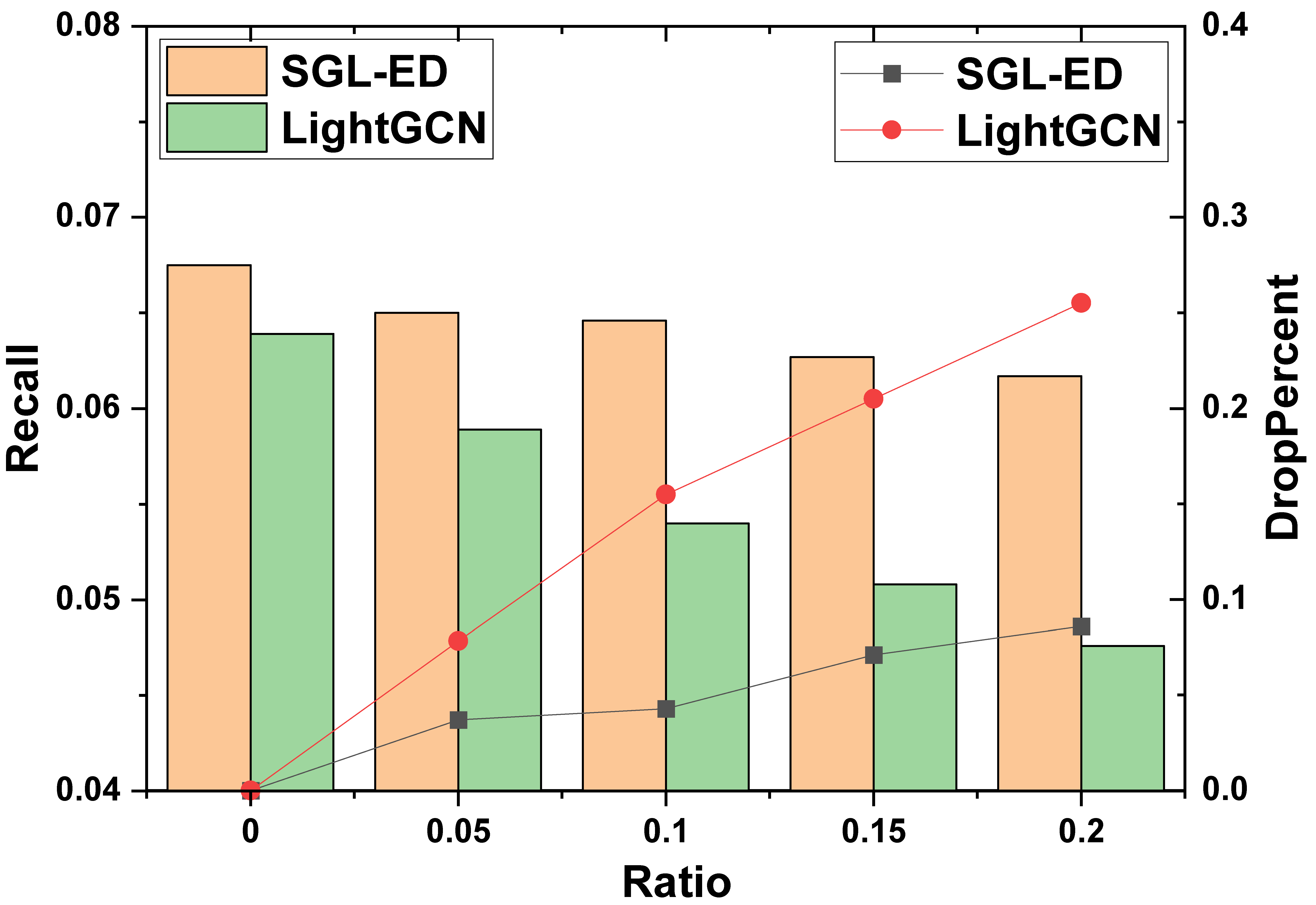}}
	\subcaptionbox{Amazon-Book\label{fig:denoise_ab}}{
		\includegraphics[width=0.23\textwidth]{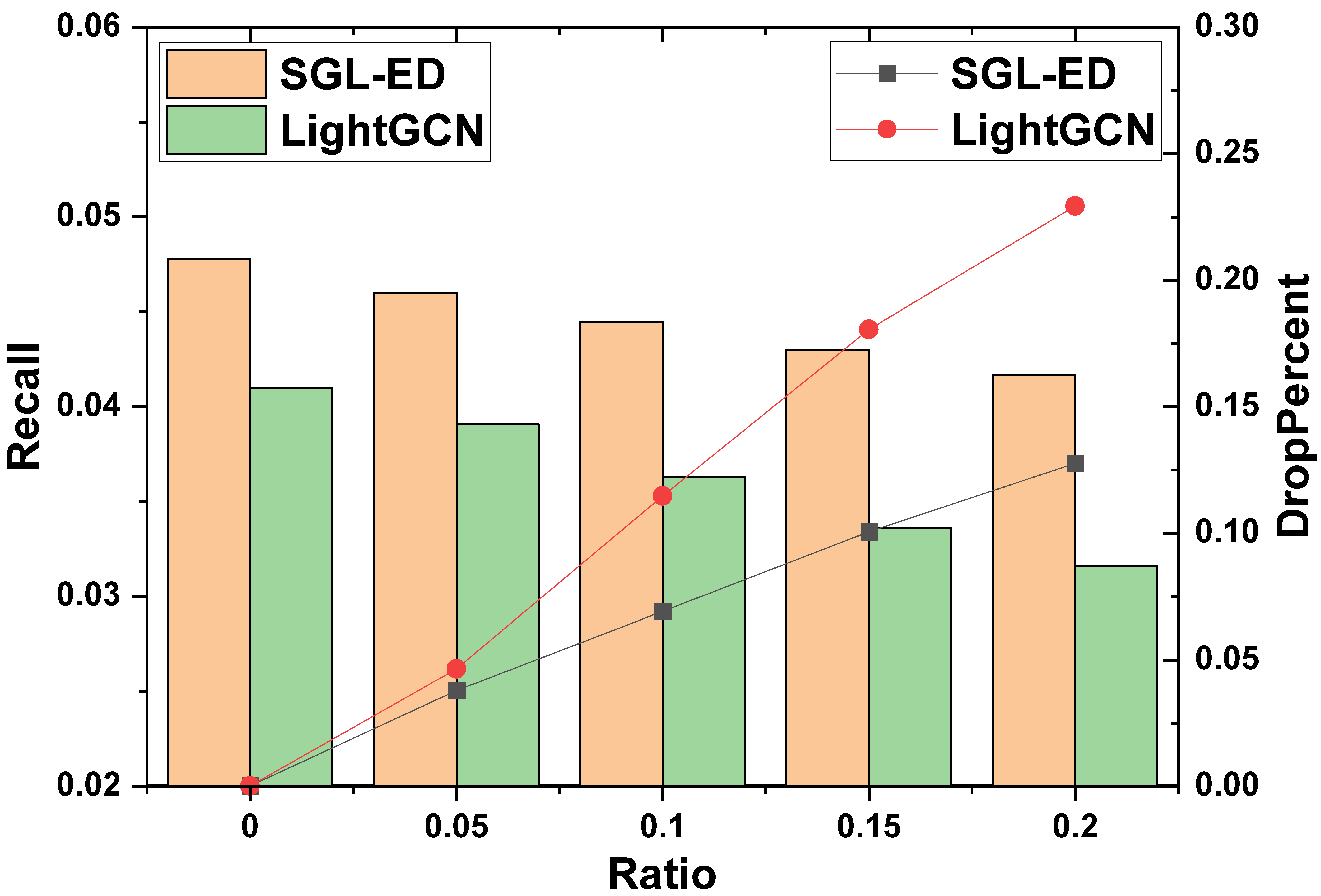}}
	\vspace{-10pt}
	\caption{Model performance \wrt noise ratio. The bar represents Recall, while the line represents the percentage of performance degradation.}
	\label{fig:denoise}
	\vspace{-10pt}
\end{figure}

We also conduct experiments to check SGL's robustness to noisy interactions.
Towards this end, we contaminate the training set by adding a certain proportion of adversarial examples (\ie 5\%, 10\%, 15\%, 20\% negative user-item interactions), while keeping the testing set unchanged.
Figure~\ref{fig:denoise} shows the results on Yelp2018 and Amazon-Book datasets.
\begin{itemize}[leftmargin=*]
    \item Clearly, adding noise data reduces the performance of SGL and LightGCN.
    However, the performance degradation of SGL is lower than that of LightGCN; moreover, as the noise ratio increases, the gap between two degradation curves becomes more apparent.
    This suggests that, by comparing differently augmented views of nodes, SGL is able to figure out useful patterns, especially informative graph structures of nodes, and reduce dependence on certain edges.
    In a nutshell, SGL offers a different angle to denoise false positive interactions in recommendation.
    \item Focusing on Amazon-Book, the performance of SGL with 20\% additional noisy interactions is still superior to LightGCN with noise-free dataset. This further justifies the superiority and robustness of SGL over LightGCN.
    \item We find that SGL is more robust on Yelp2018. The possible reason may be that Amazon-Book is much sparser than Yelp2018, and adding noisy data exerts more influence on graph structure of Amazon-Book than that of Yelp2018.
\end{itemize}

\subsection{Study of SGL (RQ3)}\label{sec:ablation_study}

\begin{table}[t]
\caption{The comparison of different SSL variants}
\vspace{-10pt}
\resizebox{0.46\textwidth}{!}{
\begin{tabular}{l|cc|cc}
\hline
\multicolumn{1}{c|}{Dataset} & \multicolumn{2}{c|}{Yelp2018} & \multicolumn{2}{c}{Amazon-Book} \\ \hline
\multicolumn{1}{c|}{Method}  & Recall    & NDCG     & Recall     & NDCG      \\ \hline\hline
SGL-ED-batch                          & 0.0670             & 0.0549            & 0.0472              & 0.0374             \\
SGL-ED-merge                          & 0.0671             & 0.0547            & 0.0464              & 0.0368             \\ \hline
SGL-pre                                & 0.0653             & 0.0533            & 0.0429              & 0.0333             \\
SGL-ED                                & \textbf{0.0675}    & \textbf{0.0555}   & \textbf{0.0478}     & \textbf{0.0379}    \\ \hline
\end{tabular}}
\label{tab:impact_ssl_objective}
\vspace{-5pt}
\end{table}

We move on to studying different designs in SGL.
We first investigate the impact of hyper-parameter $\tau$.
We then explore the potential of adopting SGL as a pretraining for existing graph-based recommendation model.
Lastly, we study the influence of negatives in the SSL objective function.
Due to the space limitation, we omit the results on iFashion which have a similar trend to that on Yelp2018 and Amazon-Book.

\subsubsection{\textbf{Effect of Temperature} $\tau$}\label{sec:exp_tau}
\begin{figure}[t]
	\centering
	\subcaptionbox{Yelp2018\label{fig:tau_yelp}}{
		\includegraphics[width=0.21\textwidth]{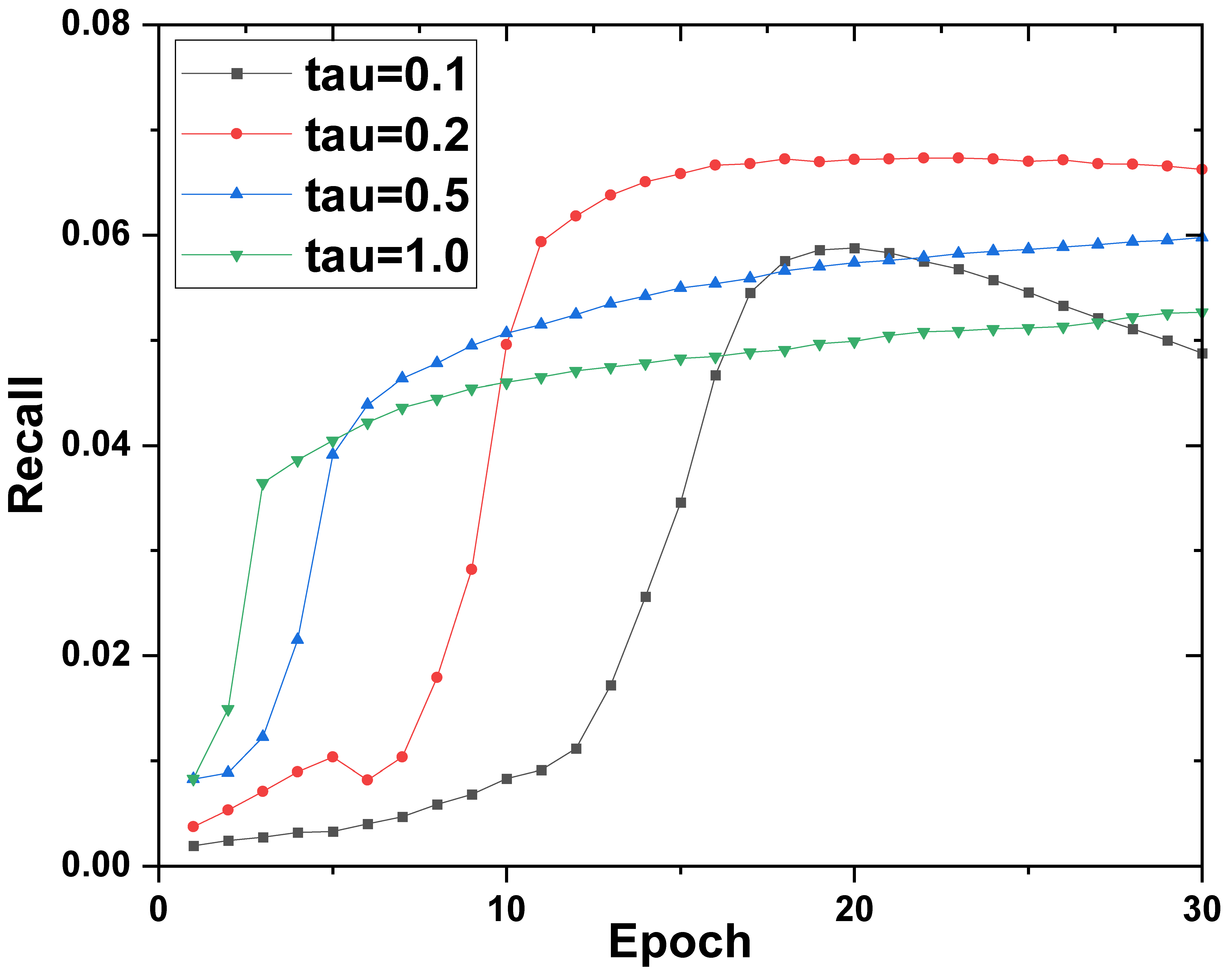}}
	\subcaptionbox{Amazon-Book\label{fig:tau_ab}}{
		\includegraphics[width=0.21\textwidth]{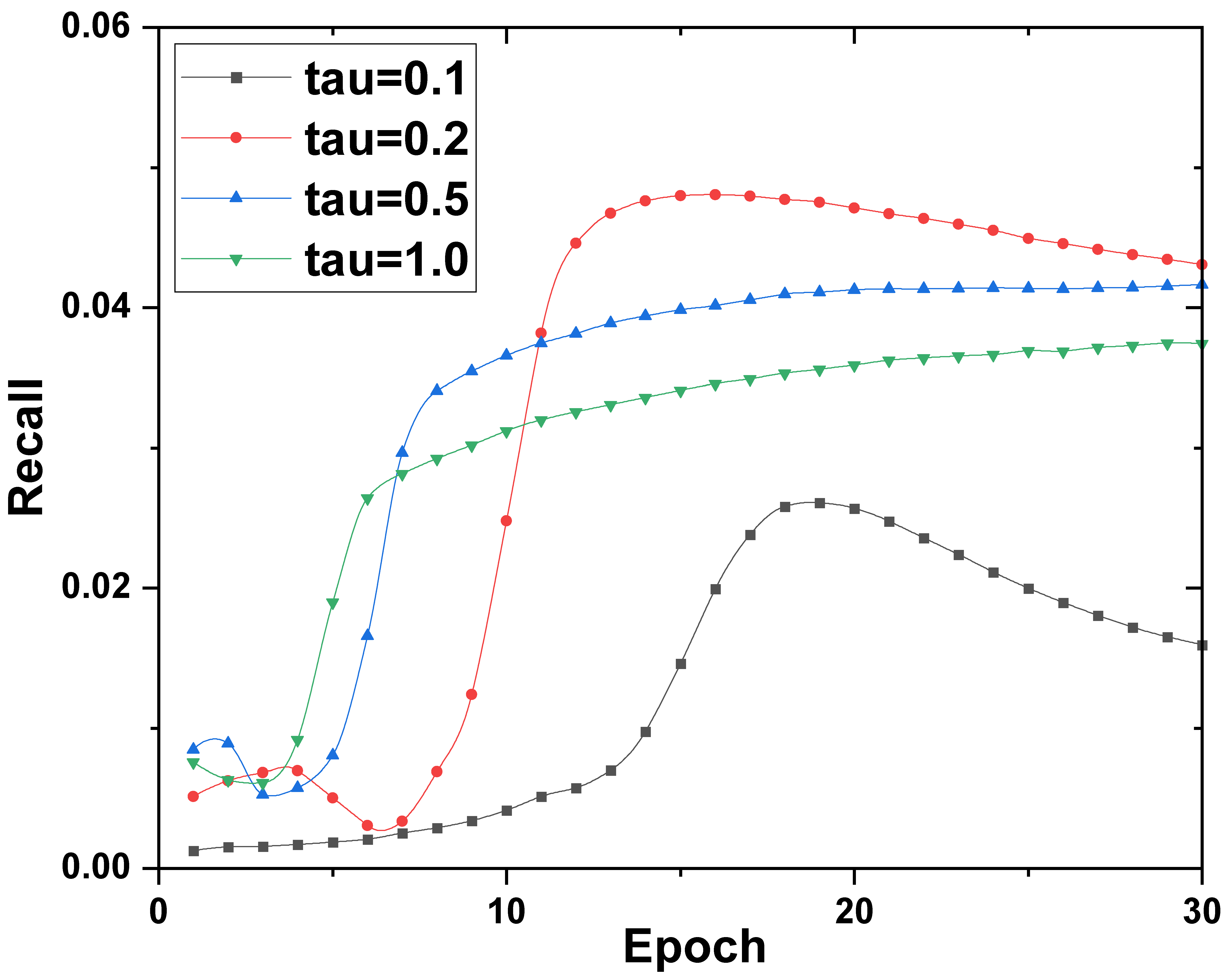}}
	\vspace{-10pt}
	\caption{Model performance as adjusting $\tau$.}
	\label{fig:study_tau}
	\vspace{-10pt}
\end{figure}
As verified in Section~\ref{sec:hard_neg_mining}, $\tau$ play a critical role in hard negative mining.
Figure~\ref{fig:study_tau} shows the curves of model performance \wrt different $\tau$. 

We can observe that:
(1) Increasing the value of $\tau$ (\eg 1.0) will lead to poorer performance and take more training epochs to converge, which fall short in the ability to distinguish hard negatives from easy negatives.
(2) In contrast, fixing $\tau$ to a too small value (\eg 0.1) will hurt the model performance, since the gradients of a few negatives dominate the optimization, losing the supremacy of adding multiple negative samples in the SSL objective.
In a nutshell, we suggest to tun $\tau$ in the range of $0.1 \sim 1.0$ carefully.

\subsubsection{\textbf{Effect of Pre-training}}\label{sec:pretrain}
The foregoing experiments have shown the effectiveness of SGL, where the main supervised task and the self-supervised task are jointly optimized.
Here we would like to answer the question: Can the recommendation performance benefit from the pre-trained model?
Towards this goal, we first pre-train the self-supervised task to obtain the model parameters, use them to initialize LightGCN, and then fine-tune the model via optimizing the main task.
We term this variant as SGL-pre and show the comparison with SGL in Table~\ref{tab:impact_ssl_objective}.
Clearly, although SGL-pre performs worse than SGL-ED on both datasets, the results of SGL-pre are still better than that of LightGCN (\cf Table~\ref{Table:comparion_with_LightGCN}).
Our self-supervised task is able to offer a better initialization for LightGCN, which is consistent to the observation in previous studies~\cite{SimCLR}.
However, the better performance of joint training admits that the representations in the main and auxiliary tasks are mutually enhanced with each other.

\subsubsection{\textbf{Effect of Negatives}}
Moreover, we also study the effect of the choice of negative samples in the auxiliary task.
Two variants are considered: (1) SGL-ED-batch, which differentiates node types and treat users and items in a mini-batch as negative views for users and items, separately, and (2) SGL-ED-merge, which treat nodes within a mini-batch as negative, without differentiating the node types.
We report the comparison in Table~\ref{tab:impact_ssl_objective}.
SGL-ED-batch performs better than SGL-ED-merge, which indicates the necessity for distinguishing types of heterogeneous nodes.
Moreover, SGL-ED-batch is on a par with SGL-ED that treats the whole spaces of users and items as negative. It suggests that training the SSL task in mini-batching is an efficient alternative.

\section{Related Work}
In this section, we separately review two tasks related to our work: graph-based recommendation and self-supervised learning.

\subsection{Graph-based Recommendation}
Previous studies on graph-based recommendation can be categorized into: \textit{model-level} and \textit{graph-level} methods, regarding their focus.
The model-level methods focus on model design for mining the user-item graph. The research attention has evolved from the random walk that encodes the graph structure as transition probabilities~\cite{DBLP:conf/www/BalujaSSJYKRA08}, to GCN that propagates user and item embeddings over the graph~\cite{GCMC,NGCF,PinSage,LightGCN}. Recently, attention mechanism is introduced into GCN-based recommendation models~\cite{DBLP:conf/ijcai/ChenGRHXGYZ19}, which learns to weigh the neighbors so as to capture the more informative user-item interactions.
A surge of attention has also been dedicated to graph-level methods, which enriches the user-item graph by accounting for side information apart from user-item interactions, which ranges from user social relations~\cite{DBLP:conf/www/BagciK16,RenLLWR17,yuan2020parameter}, item co-occurrence~\cite{DBLP:conf/www/BalujaSSJYKRA08}, to user and item attributes~\cite{Fi-GNN}.
Recently, Knowledge Graph (KG) is also unified with the user-item graph, which enables considering the detailed types of linkage between items~\cite{KGAT,KTUP,KGCN}. 

Despite the tremendous efforts devoted on these methods, all the existing work adopts the paradigm of supervised learning for model training. 
This work is in an orthogonal direction, which explores self-supervised learning, opening up a new research line of graph-based recommendation.

\subsection{Self-supervised Learning}
Studies on self-supervised learning can be roughly categorized into two branches: \textit{generative models}~\cite{word2vec,PixelCNN,DBLP:conf/naacl/DevlinCLT19} and \textit{contrastive models}~\cite{DBLP:conf/iclr/GidarisSK18,CPC, MoCo,SimCLR}. 
Auto-encoding is the most popular generative model which learns to reconstruct the input data, where noises can be intentionally added to enhance model robustness~\cite{DBLP:conf/naacl/DevlinCLT19}. 
Contrastive models learn to compare through a Noise Contrastive Estimation (NCE) objective, which can be in either global-local contrast~\cite{DeepInfoMax,CPC} or global-global contrast manner~\cite{SimCLR,MoCo}.
The former focuses on modeling the relationship between the local part of a sample and its global context representation.
While the latter directly performs comparison between different samples,
which typically requires multiple views of samples~\cite{CMC,MoCo,SimCLR}.
SSL has also been applied on graph data~\cite{icml/YouCWS20,corr/abs-2010-14945,nips/YouCSCWS20}. For instance, InfoGraph~\cite{InfoGraph} and DGI~\cite{DGI} learns node representations according to mutual information between a node and the local structure. In addition, Hu \textit{et al.} \cite{DBLP:conf/iclr/HuLGZLPL20} extend the idea to learn GCN for graph representation.
Furthermore, Kaveh \textit{et al.}~\cite{icml2020_1971} adopt the contrastive model for learning both node and graph representation, which contrasts node representations from one view with graph representation of another view.
Besides, GCC~\cite{gcc} leverages instance discrimination as the pretext task for graph structural information pre-training.
These studies focused on general graphs while leaving the inherent properties of bipartite graph unconsidered.

To the best of our knowledge, very limited works exist in combining SSL with recommendation to date. A very recent one is $S^3$-Rec~\cite{S^3-Rec} for sequential recommendation which utilizes the mutual information maximization principle to learn the correlations among attribute, item, subsequence, and sequence. Another attempt~\cite{DBLP:journals/corr/abs-2007-12865} also adopts a multi-task framework with SSL. However, it differs from our work in:
(1) \cite{DBLP:journals/corr/abs-2007-12865} uses two-tower DNN as encoder, while our work sheds lights on graph-based recommendation, and devised three augmentation operators on graph structure.
(2) \cite{DBLP:journals/corr/abs-2007-12865} utilizes categorical metadata features as model input, while our work considers a more general collaborative filtering setting with only ID as feature.
\section{Conclusion and Future Work}
In this work, we recognized the limitations of graph-based recommendation under general supervised learning paradigm and explored the potential of SSL to solve the limitations. 
In particular, we proposed a model-agnostic framework SGL to supplement the supervised recommendation task with self-supervised learning on user-item graph.  
From the perspective of graph structure, 
we devised three types of data augmentation from different aspects to construct the auxiliary contrastive task. 
We conducted extensive experiments on three benchmark datasets, justifying the advantages of our proposal regarding long-tail recommendation, training convergence and robustness against noisy interactions.

This work represents an initial attempt to exploit self-supervised learning for recommendation and opens up new research possibilities. 
In future work, we would like to make SSL more entrenched with recommendation tasks.
Going beyond the stochastic selections on graph structure, we plan to explore new perspectives, such as counterfactual learning to identify influential data points, to create more powerful data augmentations.
Moreover, we will focus on pre-training and fine-tuning in recommendation, --- that is, to pre-train a model which captures universal and transferable patterns of users across multiple domains or datasets, and fine-tune it on upcoming domains or datasets.
Another promising direction is fulfilling the potential of SSL to address the long-tail issue.
We hope the development of SGL is beneficial for improving the generalization and transferability of recommender models.  

\noindent\textbf{Acknowledgement}.
This work is supported by the National Natural Science Foundation of China (U19A2079, 61972372) and National Key Research and Development Program of China (2020AAA0106000).

\bibliographystyle{ACM-Reference-Format}
\balance
\bibliography{SGL-SIGIR2021}


\begin{thebibliography}{56}


\ifx \showCODEN    \undefined \def \showCODEN     #1{\unskip}     \fi
\ifx \showDOI      \undefined \def \showDOI       #1{#1}\fi
\ifx \showISBNx    \undefined \def \showISBNx     #1{\unskip}     \fi
\ifx \showISBNxiii \undefined \def \showISBNxiii  #1{\unskip}     \fi
\ifx \showISSN     \undefined \def \showISSN      #1{\unskip}     \fi
\ifx \showLCCN     \undefined \def \showLCCN      #1{\unskip}     \fi
\ifx \shownote     \undefined \def \shownote      #1{#1}          \fi
\ifx \showarticletitle \undefined \def \showarticletitle #1{#1}   \fi
\ifx \showURL      \undefined \def \showURL       {\relax}        \fi
\providecommand\bibfield[2]{#2}
\providecommand\bibinfo[2]{#2}
\providecommand\natexlab[1]{#1}
\providecommand\showeprint[2][]{arXiv:#2}

\bibitem[\protect\citeauthoryear{Bagci and Karagoz}{Bagci and Karagoz}{2016}]%
        {DBLP:conf/www/BagciK16}
\bibfield{author}{\bibinfo{person}{Hakan Bagci} {and} \bibinfo{person}{Pinar
  Karagoz}.} \bibinfo{year}{2016}\natexlab{}.
\newblock \showarticletitle{Context-Aware Friend Recommendation for Location
  Based Social Networks using Random Walk}. In
  \bibinfo{booktitle}{\emph{{WWW}}}. \bibinfo{pages}{531--536}.
\newblock


\bibitem[\protect\citeauthoryear{Baluja, Seth, Sivakumar, Jing, Yagnik, Kumar,
  Ravichandran, and Aly}{Baluja et~al\mbox{.}}{2008}]%
        {DBLP:conf/www/BalujaSSJYKRA08}
\bibfield{author}{\bibinfo{person}{Shumeet Baluja}, \bibinfo{person}{Rohan
  Seth}, \bibinfo{person}{D. Sivakumar}, \bibinfo{person}{Yushi Jing},
  \bibinfo{person}{Jay Yagnik}, \bibinfo{person}{Shankar Kumar},
  \bibinfo{person}{Deepak Ravichandran}, {and} \bibinfo{person}{Mohamed Aly}.}
  \bibinfo{year}{2008}\natexlab{}.
\newblock \showarticletitle{Video suggestion and discovery for youtube: taking
  random walks through the view graph}. In \bibinfo{booktitle}{\emph{{WWW}}}.
  \bibinfo{pages}{895--904}.
\newblock


\bibitem[\protect\citeauthoryear{Bayer, He, Kanagal, and Rendle}{Bayer
  et~al\mbox{.}}{2017}]%
        {iCD}
\bibfield{author}{\bibinfo{person}{Immanuel Bayer}, \bibinfo{person}{Xiangnan
  He}, \bibinfo{person}{Bhargav Kanagal}, {and} \bibinfo{person}{Steffen
  Rendle}.} \bibinfo{year}{2017}\natexlab{}.
\newblock \showarticletitle{A Generic Coordinate Descent Framework for Learning
  from Implicit Feedback}. In \bibinfo{booktitle}{\emph{{WWW}}}.
  \bibinfo{pages}{1341--1350}.
\newblock


\bibitem[\protect\citeauthoryear{Cao, Wang, He, Hu, and Chua}{Cao
  et~al\mbox{.}}{2019}]%
        {KTUP}
\bibfield{author}{\bibinfo{person}{Yixin Cao}, \bibinfo{person}{Xiang Wang},
  \bibinfo{person}{Xiangnan He}, \bibinfo{person}{Zikun Hu}, {and}
  \bibinfo{person}{Tat{-}Seng Chua}.} \bibinfo{year}{2019}\natexlab{}.
\newblock \showarticletitle{Unifying Knowledge Graph Learning and
  Recommendation: Towards a Better Understanding of User Preferences}. In
  \bibinfo{booktitle}{\emph{{WWW}}}. \bibinfo{pages}{151--161}.
\newblock


\bibitem[\protect\citeauthoryear{Chen, Dong, Wang, Feng, Wang, and He}{Chen
  et~al\mbox{.}}{2020a}]%
        {abs-2010-03240}
\bibfield{author}{\bibinfo{person}{Jiawei Chen}, \bibinfo{person}{Hande Dong},
  \bibinfo{person}{Xiang Wang}, \bibinfo{person}{Fuli Feng},
  \bibinfo{person}{Meng Wang}, {and} \bibinfo{person}{Xiangnan He}.}
  \bibinfo{year}{2020}\natexlab{a}.
\newblock \showarticletitle{Bias and Debias in Recommender System: {A} Survey
  and Future Directions}.
\newblock \bibinfo{journal}{\emph{CoRR}}  \bibinfo{volume}{abs/2010.03240}
  (\bibinfo{year}{2020}).
\newblock


\bibitem[\protect\citeauthoryear{Chen, Kornblith, Norouzi, and Hinton}{Chen
  et~al\mbox{.}}{2020b}]%
        {SimCLR}
\bibfield{author}{\bibinfo{person}{Ting Chen}, \bibinfo{person}{Simon
  Kornblith}, \bibinfo{person}{Mohammad Norouzi}, {and}
  \bibinfo{person}{Geoffrey~E. Hinton}.} \bibinfo{year}{2020}\natexlab{b}.
\newblock \showarticletitle{A Simple Framework for Contrastive Learning of
  Visual Representations}.
\newblock \bibinfo{journal}{\emph{CoRR}}  \bibinfo{volume}{abs/2002.05709}
  (\bibinfo{year}{2020}).
\newblock


\bibitem[\protect\citeauthoryear{Chen, Gu, Ren, He, Xie, Guo, Yin, and
  Zhang}{Chen et~al\mbox{.}}{2019a}]%
        {DBLP:conf/ijcai/ChenGRHXGYZ19}
\bibfield{author}{\bibinfo{person}{Weijian Chen}, \bibinfo{person}{Yulong Gu},
  \bibinfo{person}{Zhaochun Ren}, \bibinfo{person}{Xiangnan He},
  \bibinfo{person}{Hongtao Xie}, \bibinfo{person}{Tong Guo},
  \bibinfo{person}{Dawei Yin}, {and} \bibinfo{person}{Yongdong Zhang}.}
  \bibinfo{year}{2019}\natexlab{a}.
\newblock \showarticletitle{Semi-supervised User Profiling with Heterogeneous
  Graph Attention Networks}. In \bibinfo{booktitle}{\emph{IJCAI}}.
  \bibinfo{pages}{2116--2122}.
\newblock


\bibitem[\protect\citeauthoryear{Chen, Huang, Xu, Guo, Guo, Sun, Li, Pfadler,
  Zhao, and Zhao}{Chen et~al\mbox{.}}{2019b}]%
        {POG}
\bibfield{author}{\bibinfo{person}{Wen Chen}, \bibinfo{person}{Pipei Huang},
  \bibinfo{person}{Jiaming Xu}, \bibinfo{person}{Xin Guo},
  \bibinfo{person}{Cheng Guo}, \bibinfo{person}{Fei Sun}, \bibinfo{person}{Chao
  Li}, \bibinfo{person}{Andreas Pfadler}, \bibinfo{person}{Huan Zhao}, {and}
  \bibinfo{person}{Binqiang Zhao}.} \bibinfo{year}{2019}\natexlab{b}.
\newblock \showarticletitle{{POG:} Personalized Outfit Generation for Fashion
  Recommendation at Alibaba iFashion}. In \bibinfo{booktitle}{\emph{{SIGKDD}}}.
  \bibinfo{pages}{2662--2670}.
\newblock


\bibitem[\protect\citeauthoryear{Clauset, Shalizi, and Newman}{Clauset
  et~al\mbox{.}}{2009}]%
        {DBLP:journals/siamrev/ClausetSN09}
\bibfield{author}{\bibinfo{person}{Aaron Clauset},
  \bibinfo{person}{Cosma~Rohilla Shalizi}, {and} \bibinfo{person}{Mark E.~J.
  Newman}.} \bibinfo{year}{2009}\natexlab{}.
\newblock \showarticletitle{Power-Law Distributions in Empirical Data}.
\newblock \bibinfo{journal}{\emph{{SIAM}}} \bibinfo{volume}{51},
  \bibinfo{number}{4} (\bibinfo{year}{2009}), \bibinfo{pages}{661--703}.
\newblock


\bibitem[\protect\citeauthoryear{Devlin, Chang, Lee, and Toutanova}{Devlin
  et~al\mbox{.}}{2019}]%
        {DBLP:conf/naacl/DevlinCLT19}
\bibfield{author}{\bibinfo{person}{Jacob Devlin}, \bibinfo{person}{Ming{-}Wei
  Chang}, \bibinfo{person}{Kenton Lee}, {and} \bibinfo{person}{Kristina
  Toutanova}.} \bibinfo{year}{2019}\natexlab{}.
\newblock \showarticletitle{{BERT:} Pre-training of Deep Bidirectional
  Transformers for Language Understanding}. In
  \bibinfo{booktitle}{\emph{{NAACL-HLT}}}. \bibinfo{pages}{4171--4186}.
\newblock


\bibitem[\protect\citeauthoryear{Gidaris, Singh, and Komodakis}{Gidaris
  et~al\mbox{.}}{2018}]%
        {DBLP:conf/iclr/GidarisSK18}
\bibfield{author}{\bibinfo{person}{Spyros Gidaris}, \bibinfo{person}{Praveer
  Singh}, {and} \bibinfo{person}{Nikos Komodakis}.}
  \bibinfo{year}{2018}\natexlab{}.
\newblock \showarticletitle{Unsupervised Representation Learning by Predicting
  Image Rotations}. In \bibinfo{booktitle}{\emph{{ICLR}}}.
\newblock


\bibitem[\protect\citeauthoryear{Gilmer, Schoenholz, Riley, Vinyals, and
  Dahl}{Gilmer et~al\mbox{.}}{2017}]%
        {GCN}
\bibfield{author}{\bibinfo{person}{Justin Gilmer}, \bibinfo{person}{Samuel~S.
  Schoenholz}, \bibinfo{person}{Patrick~F. Riley}, \bibinfo{person}{Oriol
  Vinyals}, {and} \bibinfo{person}{George~E. Dahl}.}
  \bibinfo{year}{2017}\natexlab{}.
\newblock \showarticletitle{Neural Message Passing for Quantum Chemistry}. In
  \bibinfo{booktitle}{\emph{{ICML}}}, Vol.~\bibinfo{volume}{70}.
  \bibinfo{pages}{1263--1272}.
\newblock


\bibitem[\protect\citeauthoryear{Glorot and Bengio}{Glorot and Bengio}{2010}]%
        {Xavier}
\bibfield{author}{\bibinfo{person}{Xavier Glorot} {and} \bibinfo{person}{Yoshua
  Bengio}.} \bibinfo{year}{2010}\natexlab{}.
\newblock \showarticletitle{Understanding the difficulty of training deep
  feedforward neural networks}. In \bibinfo{booktitle}{\emph{{AISTATS}}},
  Vol.~\bibinfo{volume}{9}. \bibinfo{pages}{249--256}.
\newblock


\bibitem[\protect\citeauthoryear{Gutmann and Hyv{\"{a}}rinen}{Gutmann and
  Hyv{\"{a}}rinen}{2010}]%
        {InfoNCE}
\bibfield{author}{\bibinfo{person}{Michael Gutmann} {and} \bibinfo{person}{Aapo
  Hyv{\"{a}}rinen}.} \bibinfo{year}{2010}\natexlab{}.
\newblock \showarticletitle{Noise-contrastive estimation: {A} new estimation
  principle for unnormalized statistical models}. In
  \bibinfo{booktitle}{\emph{{AISTATS}}}, Vol.~\bibinfo{volume}{9}.
  \bibinfo{pages}{297--304}.
\newblock


\bibitem[\protect\citeauthoryear{Hamilton, Ying, and Leskovec}{Hamilton
  et~al\mbox{.}}{2017}]%
        {GraphSage}
\bibfield{author}{\bibinfo{person}{William~L. Hamilton},
  \bibinfo{person}{Zhitao Ying}, {and} \bibinfo{person}{Jure Leskovec}.}
  \bibinfo{year}{2017}\natexlab{}.
\newblock \showarticletitle{Inductive Representation Learning on Large Graphs}.
  In \bibinfo{booktitle}{\emph{{NeurIPS}}}. \bibinfo{pages}{1024--1034}.
\newblock


\bibitem[\protect\citeauthoryear{Hassani and Khasahmadi}{Hassani and
  Khasahmadi}{2020}]%
        {icml2020_1971}
\bibfield{author}{\bibinfo{person}{Kaveh Hassani} {and}
  \bibinfo{person}{Amir~Hosein Khasahmadi}.} \bibinfo{year}{2020}\natexlab{}.
\newblock \showarticletitle{Contrastive Multi-View Representation Learning on
  Graphs}.
\newblock In \bibinfo{booktitle}{\emph{ICML}}. \bibinfo{pages}{3451--3461}.
\newblock


\bibitem[\protect\citeauthoryear{He, Fan, Wu, Xie, and Girshick}{He
  et~al\mbox{.}}{2020b}]%
        {MoCo}
\bibfield{author}{\bibinfo{person}{Kaiming He}, \bibinfo{person}{Haoqi Fan},
  \bibinfo{person}{Yuxin Wu}, \bibinfo{person}{Saining Xie}, {and}
  \bibinfo{person}{Ross~B. Girshick}.} \bibinfo{year}{2020}\natexlab{b}.
\newblock \showarticletitle{Momentum Contrast for Unsupervised Visual
  Representation Learning}. In \bibinfo{booktitle}{\emph{{CVPR}}}.
  \bibinfo{pages}{9726--9735}.
\newblock


\bibitem[\protect\citeauthoryear{He and McAuley}{He and McAuley}{2016}]%
        {DBLP:conf/www/HeM16}
\bibfield{author}{\bibinfo{person}{Ruining He} {and} \bibinfo{person}{Julian~J.
  McAuley}.} \bibinfo{year}{2016}\natexlab{}.
\newblock \showarticletitle{Ups and Downs: Modeling the Visual Evolution of
  Fashion Trends with One-Class Collaborative Filtering}. In
  \bibinfo{booktitle}{\emph{{WWW}}}. \bibinfo{pages}{507--517}.
\newblock


\bibitem[\protect\citeauthoryear{He, Deng, Wang, Li, Zhang, and Wang}{He
  et~al\mbox{.}}{2020a}]%
        {LightGCN}
\bibfield{author}{\bibinfo{person}{Xiangnan He}, \bibinfo{person}{Kuan Deng},
  \bibinfo{person}{Xiang Wang}, \bibinfo{person}{Yan Li},
  \bibinfo{person}{Yong{-}Dong Zhang}, {and} \bibinfo{person}{Meng Wang}.}
  \bibinfo{year}{2020}\natexlab{a}.
\newblock \showarticletitle{LightGCN: Simplifying and Powering Graph
  Convolution Network for Recommendation}. In
  \bibinfo{booktitle}{\emph{{SIGIR}}}. \bibinfo{pages}{639--648}.
\newblock


\bibitem[\protect\citeauthoryear{He, He, Song, Liu, Jiang, and Chua}{He
  et~al\mbox{.}}{2018}]%
        {NAIS}
\bibfield{author}{\bibinfo{person}{Xiangnan He}, \bibinfo{person}{Zhankui He},
  \bibinfo{person}{Jingkuan Song}, \bibinfo{person}{Zhenguang Liu},
  \bibinfo{person}{Yu{-}Gang Jiang}, {and} \bibinfo{person}{Tat{-}Seng Chua}.}
  \bibinfo{year}{2018}\natexlab{}.
\newblock \showarticletitle{{NAIS:} Neural Attentive Item Similarity Model for
  Recommendation}.
\newblock \bibinfo{journal}{\emph{TKDE}} \bibinfo{volume}{30},
  \bibinfo{number}{12} (\bibinfo{year}{2018}), \bibinfo{pages}{2354--2366}.
\newblock


\bibitem[\protect\citeauthoryear{He, Liao, Zhang, Nie, Hu, and Chua}{He
  et~al\mbox{.}}{2017}]%
        {NeuMF}
\bibfield{author}{\bibinfo{person}{Xiangnan He}, \bibinfo{person}{Lizi Liao},
  \bibinfo{person}{Hanwang Zhang}, \bibinfo{person}{Liqiang Nie},
  \bibinfo{person}{Xia Hu}, {and} \bibinfo{person}{Tat{-}Seng Chua}.}
  \bibinfo{year}{2017}\natexlab{}.
\newblock \showarticletitle{Neural Collaborative Filtering}. In
  \bibinfo{booktitle}{\emph{{WWW}}}. \bibinfo{pages}{173--182}.
\newblock


\bibitem[\protect\citeauthoryear{Hjelm, Fedorov, Lavoie{-}Marchildon, Grewal,
  Bachman, Trischler, and Bengio}{Hjelm et~al\mbox{.}}{2019}]%
        {DeepInfoMax}
\bibfield{author}{\bibinfo{person}{R.~Devon Hjelm}, \bibinfo{person}{Alex
  Fedorov}, \bibinfo{person}{Samuel Lavoie{-}Marchildon},
  \bibinfo{person}{Karan Grewal}, \bibinfo{person}{Philip Bachman},
  \bibinfo{person}{Adam Trischler}, {and} \bibinfo{person}{Yoshua Bengio}.}
  \bibinfo{year}{2019}\natexlab{}.
\newblock \showarticletitle{Learning deep representations by mutual information
  estimation and maximization}. In \bibinfo{booktitle}{\emph{{ICLR}}}.
\newblock


\bibitem[\protect\citeauthoryear{Hu, Liu, Gomes, Zitnik, Liang, Pande, and
  Leskovec}{Hu et~al\mbox{.}}{2020}]%
        {DBLP:conf/iclr/HuLGZLPL20}
\bibfield{author}{\bibinfo{person}{Weihua Hu}, \bibinfo{person}{Bowen Liu},
  \bibinfo{person}{Joseph Gomes}, \bibinfo{person}{Marinka Zitnik},
  \bibinfo{person}{Percy Liang}, \bibinfo{person}{Vijay~S. Pande}, {and}
  \bibinfo{person}{Jure Leskovec}.} \bibinfo{year}{2020}\natexlab{}.
\newblock \showarticletitle{Strategies for Pre-training Graph Neural Networks}.
  In \bibinfo{booktitle}{\emph{{ICLR}}}.
\newblock


\bibitem[\protect\citeauthoryear{Khosla, Teterwak, Wang, Sarna, Tian, Isola,
  Maschinot, Liu, and Krishnan}{Khosla et~al\mbox{.}}{2020}]%
        {KhoslaTWSTIMLK20}
\bibfield{author}{\bibinfo{person}{Prannay Khosla}, \bibinfo{person}{Piotr
  Teterwak}, \bibinfo{person}{Chen Wang}, \bibinfo{person}{Aaron Sarna},
  \bibinfo{person}{Yonglong Tian}, \bibinfo{person}{Phillip Isola},
  \bibinfo{person}{Aaron Maschinot}, \bibinfo{person}{Ce Liu}, {and}
  \bibinfo{person}{Dilip Krishnan}.} \bibinfo{year}{2020}\natexlab{}.
\newblock \showarticletitle{Supervised Contrastive Learning}. In
  \bibinfo{booktitle}{\emph{NeurIPS}}.
\newblock


\bibitem[\protect\citeauthoryear{Koren}{Koren}{2008}]%
        {SVD++}
\bibfield{author}{\bibinfo{person}{Yehuda Koren}.}
  \bibinfo{year}{2008}\natexlab{}.
\newblock \showarticletitle{Factorization meets the neighborhood: a
  multifaceted collaborative filtering model}. In
  \bibinfo{booktitle}{\emph{{KDD}}}. \bibinfo{pages}{426--434}.
\newblock


\bibitem[\protect\citeauthoryear{Lan, Chen, Goodman, Gimpel, Sharma, and
  Soricut}{Lan et~al\mbox{.}}{2020}]%
        {DBLP:conf/iclr/LanCGGSS20}
\bibfield{author}{\bibinfo{person}{Zhenzhong Lan}, \bibinfo{person}{Mingda
  Chen}, \bibinfo{person}{Sebastian Goodman}, \bibinfo{person}{Kevin Gimpel},
  \bibinfo{person}{Piyush Sharma}, {and} \bibinfo{person}{Radu Soricut}.}
  \bibinfo{year}{2020}\natexlab{}.
\newblock \showarticletitle{{ALBERT:} {A} Lite {BERT} for Self-supervised
  Learning of Language Representations}. In \bibinfo{booktitle}{\emph{{ICLR}}}.
\newblock


\bibitem[\protect\citeauthoryear{Li, Cui, Wu, Zhang, and Wang}{Li
  et~al\mbox{.}}{2019}]%
        {Fi-GNN}
\bibfield{author}{\bibinfo{person}{Zekun Li}, \bibinfo{person}{Zeyu Cui},
  \bibinfo{person}{Shu Wu}, \bibinfo{person}{Xiaoyu Zhang}, {and}
  \bibinfo{person}{Liang Wang}.} \bibinfo{year}{2019}\natexlab{}.
\newblock \showarticletitle{Fi-GNN: Modeling Feature Interactions via Graph
  Neural Networks for {CTR} Prediction}. In \bibinfo{booktitle}{\emph{{CIKM}}}.
  \bibinfo{pages}{539--548}.
\newblock


\bibitem[\protect\citeauthoryear{Liang, Krishnan, Hoffman, and Jebara}{Liang
  et~al\mbox{.}}{2018}]%
        {MultiVAE}
\bibfield{author}{\bibinfo{person}{Dawen Liang}, \bibinfo{person}{Rahul~G.
  Krishnan}, \bibinfo{person}{Matthew~D. Hoffman}, {and} \bibinfo{person}{Tony
  Jebara}.} \bibinfo{year}{2018}\natexlab{}.
\newblock \showarticletitle{Variational Autoencoders for Collaborative
  Filtering}. In \bibinfo{booktitle}{\emph{{WWW}}}. \bibinfo{pages}{689--698}.
\newblock


\bibitem[\protect\citeauthoryear{Mikolov, Sutskever, Chen, Corrado, and
  Dean}{Mikolov et~al\mbox{.}}{2013}]%
        {word2vec}
\bibfield{author}{\bibinfo{person}{Tomas Mikolov}, \bibinfo{person}{Ilya
  Sutskever}, \bibinfo{person}{Kai Chen}, \bibinfo{person}{Gregory~S. Corrado},
  {and} \bibinfo{person}{Jeffrey Dean}.} \bibinfo{year}{2013}\natexlab{}.
\newblock \showarticletitle{Distributed Representations of Words and Phrases
  and their Compositionality}. In \bibinfo{booktitle}{\emph{NIPS}}.
  \bibinfo{pages}{3111--3119}.
\newblock


\bibitem[\protect\citeauthoryear{Milojevic}{Milojevic}{2010}]%
        {PowerLaw}
\bibfield{author}{\bibinfo{person}{Stasa Milojevic}.}
  \bibinfo{year}{2010}\natexlab{}.
\newblock \showarticletitle{Power law distributions in information science:
  Making the case for logarithmic binning}.
\newblock \bibinfo{journal}{\emph{J. Assoc. Inf. Sci. Technol.}}
  \bibinfo{volume}{61}, \bibinfo{number}{12} (\bibinfo{year}{2010}),
  \bibinfo{pages}{2417--2425}.
\newblock


\bibitem[\protect\citeauthoryear{Qiu, Chen, Dong, Zhang, Yang, Ding, Wang, and
  Tang}{Qiu et~al\mbox{.}}{2020}]%
        {gcc}
\bibfield{author}{\bibinfo{person}{Jiezhong Qiu}, \bibinfo{person}{Qibin Chen},
  \bibinfo{person}{Yuxiao Dong}, \bibinfo{person}{Jing Zhang},
  \bibinfo{person}{Hongxia Yang}, \bibinfo{person}{Ming Ding},
  \bibinfo{person}{Kuansan Wang}, {and} \bibinfo{person}{Jie Tang}.}
  \bibinfo{year}{2020}\natexlab{}.
\newblock \showarticletitle{{GCC:} Graph Contrastive Coding for Graph Neural
  Network Pre-Training}. In \bibinfo{booktitle}{\emph{{KDD}}}.
  \bibinfo{pages}{1150--1160}.
\newblock


\bibitem[\protect\citeauthoryear{Ren, Liang, Li, Wang, and de~Rijke}{Ren
  et~al\mbox{.}}{2017}]%
        {RenLLWR17}
\bibfield{author}{\bibinfo{person}{Zhaochun Ren}, \bibinfo{person}{Shangsong
  Liang}, \bibinfo{person}{Piji Li}, \bibinfo{person}{Shuaiqiang Wang}, {and}
  \bibinfo{person}{Maarten de Rijke}.} \bibinfo{year}{2017}\natexlab{}.
\newblock \showarticletitle{Social Collaborative Viewpoint Regression with
  Explainable Recommendations}. In \bibinfo{booktitle}{\emph{WSDM}}.
  \bibinfo{pages}{485--494}.
\newblock


\bibitem[\protect\citeauthoryear{Rendle and Freudenthaler}{Rendle and
  Freudenthaler}{2014}]%
        {DBLP:conf/wsdm/RendleF14}
\bibfield{author}{\bibinfo{person}{Steffen Rendle} {and}
  \bibinfo{person}{Christoph Freudenthaler}.} \bibinfo{year}{2014}\natexlab{}.
\newblock \showarticletitle{Improving pairwise learning for item recommendation
  from implicit feedback}. In \bibinfo{booktitle}{\emph{{WSDM}}}.
  \bibinfo{pages}{273--282}.
\newblock


\bibitem[\protect\citeauthoryear{Rendle, Freudenthaler, Gantner, and
  Schmidt{-}Thieme}{Rendle et~al\mbox{.}}{2009}]%
        {BPR}
\bibfield{author}{\bibinfo{person}{Steffen Rendle}, \bibinfo{person}{Christoph
  Freudenthaler}, \bibinfo{person}{Zeno Gantner}, {and} \bibinfo{person}{Lars
  Schmidt{-}Thieme}.} \bibinfo{year}{2009}\natexlab{}.
\newblock \showarticletitle{{BPR:} Bayesian Personalized Ranking from Implicit
  Feedback}. In \bibinfo{booktitle}{\emph{{UAI}}}. \bibinfo{pages}{452--461}.
\newblock


\bibitem[\protect\citeauthoryear{Sun, Hoffmann, Verma, and Tang}{Sun
  et~al\mbox{.}}{2020}]%
        {InfoGraph}
\bibfield{author}{\bibinfo{person}{Fan{-}Yun Sun}, \bibinfo{person}{Jordan
  Hoffmann}, \bibinfo{person}{Vikas Verma}, {and} \bibinfo{person}{Jian Tang}.}
  \bibinfo{year}{2020}\natexlab{}.
\newblock \showarticletitle{InfoGraph: Unsupervised and Semi-supervised
  Graph-Level Representation Learning via Mutual Information Maximization}. In
  \bibinfo{booktitle}{\emph{{ICLR}}}.
\newblock


\bibitem[\protect\citeauthoryear{Tang, Yao, Sun, Wang, Tang, Aggarwal, Mitra,
  and Wang}{Tang et~al\mbox{.}}{2020}]%
        {tang2020investigating}
\bibfield{author}{\bibinfo{person}{Xianfeng Tang}, \bibinfo{person}{Huaxiu
  Yao}, \bibinfo{person}{Yiwei Sun}, \bibinfo{person}{Yiqi Wang},
  \bibinfo{person}{Jiliang Tang}, \bibinfo{person}{Charu Aggarwal},
  \bibinfo{person}{Prasenjit Mitra}, {and} \bibinfo{person}{Suhang Wang}.}
  \bibinfo{year}{2020}\natexlab{}.
\newblock \showarticletitle{Investigating and Mitigating Degree-Related Biases
  in Graph Convolutional Networks}. In \bibinfo{booktitle}{\emph{{CIKM}}}.
\newblock


\bibitem[\protect\citeauthoryear{Tian, Krishnan, and Isola}{Tian
  et~al\mbox{.}}{2019}]%
        {CMC}
\bibfield{author}{\bibinfo{person}{Yonglong Tian}, \bibinfo{person}{Dilip
  Krishnan}, {and} \bibinfo{person}{Phillip Isola}.}
  \bibinfo{year}{2019}\natexlab{}.
\newblock \showarticletitle{Contrastive Multiview Coding}.
\newblock \bibinfo{journal}{\emph{CoRR}}  \bibinfo{volume}{abs/1906.05849}
  (\bibinfo{year}{2019}).
\newblock


\bibitem[\protect\citeauthoryear{van~den Berg, Kipf, and Welling}{van~den Berg
  et~al\mbox{.}}{2017}]%
        {GCMC}
\bibfield{author}{\bibinfo{person}{Rianne van~den Berg},
  \bibinfo{person}{Thomas~N. Kipf}, {and} \bibinfo{person}{Max Welling}.}
  \bibinfo{year}{2017}\natexlab{}.
\newblock \showarticletitle{Graph Convolutional Matrix Completion}.
\newblock \bibinfo{journal}{\emph{CoRR}}  \bibinfo{volume}{abs/1706.02263}
  (\bibinfo{year}{2017}).
\newblock


\bibitem[\protect\citeauthoryear{van~den Oord, Kalchbrenner, Espeholt,
  Kavukcuoglu, Vinyals, and Graves}{van~den Oord et~al\mbox{.}}{2016}]%
        {PixelCNN}
\bibfield{author}{\bibinfo{person}{A{\"{a}}ron van~den Oord},
  \bibinfo{person}{Nal Kalchbrenner}, \bibinfo{person}{Lasse Espeholt},
  \bibinfo{person}{Koray Kavukcuoglu}, \bibinfo{person}{Oriol Vinyals}, {and}
  \bibinfo{person}{Alex Graves}.} \bibinfo{year}{2016}\natexlab{}.
\newblock \showarticletitle{Conditional Image Generation with PixelCNN
  Decoders}. In \bibinfo{booktitle}{\emph{NIPS}}. \bibinfo{pages}{4790--4798}.
\newblock


\bibitem[\protect\citeauthoryear{van~den Oord, Li, and Vinyals}{van~den Oord
  et~al\mbox{.}}{2018}]%
        {CPC}
\bibfield{author}{\bibinfo{person}{A{\"{a}}ron van~den Oord},
  \bibinfo{person}{Yazhe Li}, {and} \bibinfo{person}{Oriol Vinyals}.}
  \bibinfo{year}{2018}\natexlab{}.
\newblock \showarticletitle{Representation Learning with Contrastive Predictive
  Coding}.
\newblock \bibinfo{journal}{\emph{CoRR}}  \bibinfo{volume}{abs/1807.03748}
  (\bibinfo{year}{2018}).
\newblock


\bibitem[\protect\citeauthoryear{Velickovic, Cucurull, Casanova, Romero,
  Li{\`{o}}, and Bengio}{Velickovic et~al\mbox{.}}{2018}]%
        {GAT}
\bibfield{author}{\bibinfo{person}{Petar Velickovic}, \bibinfo{person}{Guillem
  Cucurull}, \bibinfo{person}{Arantxa Casanova}, \bibinfo{person}{Adriana
  Romero}, \bibinfo{person}{Pietro Li{\`{o}}}, {and} \bibinfo{person}{Yoshua
  Bengio}.} \bibinfo{year}{2018}\natexlab{}.
\newblock \showarticletitle{Graph Attention Networks}. In
  \bibinfo{booktitle}{\emph{{ICLR}}}.
\newblock


\bibitem[\protect\citeauthoryear{Velickovic, Fedus, Hamilton, Li{\`{o}},
  Bengio, and Hjelm}{Velickovic et~al\mbox{.}}{2019}]%
        {DGI}
\bibfield{author}{\bibinfo{person}{Petar Velickovic}, \bibinfo{person}{William
  Fedus}, \bibinfo{person}{William~L. Hamilton}, \bibinfo{person}{Pietro
  Li{\`{o}}}, \bibinfo{person}{Yoshua Bengio}, {and} \bibinfo{person}{R.~Devon
  Hjelm}.} \bibinfo{year}{2019}\natexlab{}.
\newblock \showarticletitle{Deep Graph Infomax}. In
  \bibinfo{booktitle}{\emph{{ICLR}}}.
\newblock


\bibitem[\protect\citeauthoryear{Wang, Zhao, Xie, Li, and Guo}{Wang
  et~al\mbox{.}}{2019c}]%
        {KGCN}
\bibfield{author}{\bibinfo{person}{Hongwei Wang}, \bibinfo{person}{Miao Zhao},
  \bibinfo{person}{Xing Xie}, \bibinfo{person}{Wenjie Li}, {and}
  \bibinfo{person}{Minyi Guo}.} \bibinfo{year}{2019}\natexlab{c}.
\newblock \showarticletitle{Knowledge Graph Convolutional Networks for
  Recommender Systems}. In \bibinfo{booktitle}{\emph{{WWW}}}.
  \bibinfo{pages}{3307--3313}.
\newblock


\bibitem[\protect\citeauthoryear{Wang, Feng, He, Nie, and Chua}{Wang
  et~al\mbox{.}}{2021}]%
        {abs-2006-04153}
\bibfield{author}{\bibinfo{person}{Wenjie Wang}, \bibinfo{person}{Fuli Feng},
  \bibinfo{person}{Xiangnan He}, \bibinfo{person}{Liqiang Nie}, {and}
  \bibinfo{person}{Tat{-}Seng Chua}.} \bibinfo{year}{2021}\natexlab{}.
\newblock \showarticletitle{Denoising Implicit Feedback for Recommendation}. In
  \bibinfo{booktitle}{\emph{WSDM}}.
\newblock


\bibitem[\protect\citeauthoryear{Wang, He, Cao, Liu, and Chua}{Wang
  et~al\mbox{.}}{2019a}]%
        {KGAT}
\bibfield{author}{\bibinfo{person}{Xiang Wang}, \bibinfo{person}{Xiangnan He},
  \bibinfo{person}{Yixin Cao}, \bibinfo{person}{Meng Liu}, {and}
  \bibinfo{person}{Tat{-}Seng Chua}.} \bibinfo{year}{2019}\natexlab{a}.
\newblock \showarticletitle{{KGAT:} Knowledge Graph Attention Network for
  Recommendation}. In \bibinfo{booktitle}{\emph{SIGKDD}}.
  \bibinfo{pages}{950--958}.
\newblock


\bibitem[\protect\citeauthoryear{Wang, He, Wang, Feng, and Chua}{Wang
  et~al\mbox{.}}{2019b}]%
        {NGCF}
\bibfield{author}{\bibinfo{person}{Xiang Wang}, \bibinfo{person}{Xiangnan He},
  \bibinfo{person}{Meng Wang}, \bibinfo{person}{Fuli Feng}, {and}
  \bibinfo{person}{Tat{-}Seng Chua}.} \bibinfo{year}{2019}\natexlab{b}.
\newblock \showarticletitle{Neural Graph Collaborative Filtering}. In
  \bibinfo{booktitle}{\emph{{SIGIR}}}. \bibinfo{pages}{165--174}.
\newblock


\bibitem[\protect\citeauthoryear{Wu, Xiong, Yu, and Lin}{Wu
  et~al\mbox{.}}{2018}]%
        {ContrastiveLearning}
\bibfield{author}{\bibinfo{person}{Zhirong Wu}, \bibinfo{person}{Yuanjun
  Xiong}, \bibinfo{person}{Stella~X. Yu}, {and} \bibinfo{person}{Dahua Lin}.}
  \bibinfo{year}{2018}\natexlab{}.
\newblock \showarticletitle{Unsupervised Feature Learning via Non-Parametric
  Instance Discrimination}. In \bibinfo{booktitle}{\emph{{CVPR}}}.
  \bibinfo{pages}{3733--3742}.
\newblock


\bibitem[\protect\citeauthoryear{Xu, Hu, Leskovec, and Jegelka}{Xu
  et~al\mbox{.}}{2019}]%
        {GIN}
\bibfield{author}{\bibinfo{person}{Keyulu Xu}, \bibinfo{person}{Weihua Hu},
  \bibinfo{person}{Jure Leskovec}, {and} \bibinfo{person}{Stefanie Jegelka}.}
  \bibinfo{year}{2019}\natexlab{}.
\newblock \showarticletitle{How Powerful are Graph Neural Networks?}. In
  \bibinfo{booktitle}{\emph{{ICLR}}}.
\newblock


\bibitem[\protect\citeauthoryear{Yao, Yi, Cheng, Yu, Menon, Hong, Chi, Tjoa,
  Kang, and Ettinger}{Yao et~al\mbox{.}}{2020}]%
        {DBLP:journals/corr/abs-2007-12865}
\bibfield{author}{\bibinfo{person}{Tiansheng Yao}, \bibinfo{person}{Xinyang
  Yi}, \bibinfo{person}{Derek~Zhiyuan Cheng}, \bibinfo{person}{Felix~X. Yu},
  \bibinfo{person}{Aditya~Krishna Menon}, \bibinfo{person}{Lichan Hong},
  \bibinfo{person}{Ed~H. Chi}, \bibinfo{person}{Steve Tjoa},
  \bibinfo{person}{Jieqi Kang}, {and} \bibinfo{person}{Evan Ettinger}.}
  \bibinfo{year}{2020}\natexlab{}.
\newblock \showarticletitle{Self-supervised Learning for Deep Models in
  Recommendations}.
\newblock \bibinfo{journal}{\emph{CoRR}}  \bibinfo{volume}{abs/2007.12865}
  (\bibinfo{year}{2020}).
\newblock


\bibitem[\protect\citeauthoryear{Ying, He, Chen, Eksombatchai, Hamilton, and
  Leskovec}{Ying et~al\mbox{.}}{2018}]%
        {PinSage}
\bibfield{author}{\bibinfo{person}{Rex Ying}, \bibinfo{person}{Ruining He},
  \bibinfo{person}{Kaifeng Chen}, \bibinfo{person}{Pong Eksombatchai},
  \bibinfo{person}{William~L. Hamilton}, {and} \bibinfo{person}{Jure
  Leskovec}.} \bibinfo{year}{2018}\natexlab{}.
\newblock \showarticletitle{Graph Convolutional Neural Networks for Web-Scale
  Recommender Systems}. In \bibinfo{booktitle}{\emph{{KDD}}}.
  \bibinfo{pages}{974--983}.
\newblock


\bibitem[\protect\citeauthoryear{You, Chen, Sui, Chen, Wang, and Shen}{You
  et~al\mbox{.}}{2020a}]%
        {nips/YouCSCWS20}
\bibfield{author}{\bibinfo{person}{Yuning You}, \bibinfo{person}{Tianlong
  Chen}, \bibinfo{person}{Yongduo Sui}, \bibinfo{person}{Ting Chen},
  \bibinfo{person}{Zhangyang Wang}, {and} \bibinfo{person}{Yang Shen}.}
  \bibinfo{year}{2020}\natexlab{a}.
\newblock \showarticletitle{Graph Contrastive Learning with Augmentations}. In
  \bibinfo{booktitle}{\emph{NeurIPS}}.
\newblock


\bibitem[\protect\citeauthoryear{You, Chen, Wang, and Shen}{You
  et~al\mbox{.}}{2020b}]%
        {icml/YouCWS20}
\bibfield{author}{\bibinfo{person}{Yuning You}, \bibinfo{person}{Tianlong
  Chen}, \bibinfo{person}{Zhangyang Wang}, {and} \bibinfo{person}{Yang Shen}.}
  \bibinfo{year}{2020}\natexlab{b}.
\newblock \showarticletitle{When Does Self-Supervision Help Graph Convolutional
  Networks?}. In \bibinfo{booktitle}{\emph{{ICML}}},
  Vol.~\bibinfo{volume}{119}. \bibinfo{pages}{10871--10880}.
\newblock


\bibitem[\protect\citeauthoryear{Yuan, He, Karatzoglou, and Zhang}{Yuan
  et~al\mbox{.}}{2020}]%
        {yuan2020parameter}
\bibfield{author}{\bibinfo{person}{Fajie Yuan}, \bibinfo{person}{Xiangnan He},
  \bibinfo{person}{Alexandros Karatzoglou}, {and} \bibinfo{person}{Liguang
  Zhang}.} \bibinfo{year}{2020}\natexlab{}.
\newblock \showarticletitle{Parameter-efficient transfer from sequential
  behaviors for user modeling and recommendation}. In
  \bibinfo{booktitle}{\emph{SIGIR}}. \bibinfo{pages}{1469--1478}.
\newblock


\bibitem[\protect\citeauthoryear{Zhang, Chen, Wang, and Yu}{Zhang
  et~al\mbox{.}}{2013}]%
        {DNS}
\bibfield{author}{\bibinfo{person}{Weinan Zhang}, \bibinfo{person}{Tianqi
  Chen}, \bibinfo{person}{Jun Wang}, {and} \bibinfo{person}{Yong Yu}.}
  \bibinfo{year}{2013}\natexlab{}.
\newblock \showarticletitle{Optimizing top-n collaborative filtering via
  dynamic negative item sampling}. In \bibinfo{booktitle}{\emph{{SIGIR}}}.
  \bibinfo{pages}{785--788}.
\newblock


\bibitem[\protect\citeauthoryear{Zhou, Wang, Zhao, Zhu, Wang, Zhang, Wang, and
  Wen}{Zhou et~al\mbox{.}}{2020}]%
        {S^3-Rec}
\bibfield{author}{\bibinfo{person}{Kun Zhou}, \bibinfo{person}{Hui Wang},
  \bibinfo{person}{Wayne~Xin Zhao}, \bibinfo{person}{Yutao Zhu},
  \bibinfo{person}{Sirui Wang}, \bibinfo{person}{Fuzheng Zhang},
  \bibinfo{person}{Zhongyuan Wang}, {and} \bibinfo{person}{Ji{-}Rong Wen}.}
  \bibinfo{year}{2020}\natexlab{}.
\newblock \showarticletitle{S{\^{}}3-Rec: Self-Supervised Learning for
  Sequential Recommendation with Mutual Information Maximization}. In
  \bibinfo{booktitle}{\emph{CIKM}}.
\newblock


\bibitem[\protect\citeauthoryear{Zhu, Xu, Yu, Liu, Wu, and Wang}{Zhu
  et~al\mbox{.}}{2020}]%
        {corr/abs-2010-14945}
\bibfield{author}{\bibinfo{person}{Yanqiao Zhu}, \bibinfo{person}{Yichen Xu},
  \bibinfo{person}{Feng Yu}, \bibinfo{person}{Qiang Liu}, \bibinfo{person}{Shu
  Wu}, {and} \bibinfo{person}{Liang Wang}.} \bibinfo{year}{2020}\natexlab{}.
\newblock \showarticletitle{Graph Contrastive Learning with Adaptive
  Augmentation}.
\newblock \bibinfo{journal}{\emph{CoRR}}  \bibinfo{volume}{abs/2010.14945}
  (\bibinfo{year}{2020}).
\newblock


\end{thebibliography}

\clearpage
\appendix
\section{Gradient of InfoNCE Loss \wrt node representation}

Here we derive the Equation~\eqref{Eq:gradients}.
Since we adopt cosine similarity function in the contrastive loss, we here decouple the normalization to show its superiority inspired from~\cite{KhoslaTWSTIMLK20}.
For a user $u \in \mathcal{U}$, the contrastive loss is as follows:
\begin{gather}
    \Lapl_{ssl}^{user}(u) = -\log \frac{\exp ({s_u^{'}}^T s_u^{''}/\tau)}{\sum_{v\in \mathcal{U}} \exp ({s_u^{'}}^T s_v^{''}/\tau)}
\end{gather}
where $s_u^{'}$ and $s_u^{''}$ are the normalized representations, \ie $s_u^{'}=\frac{z_u^{'}}{\left \| z_u^{'} \right \|}, s_u^{''}=\frac{z_u^{''}}{\left \| z_u^{''} \right \|}$. The gradient of the contrastive Loss \wrt $z_u^{'}$ can be calculated via the chain rule of scalar-by-vector:
\begin{gather}
    \frac{\partial \Lapl_{ssl}^{user}(u)}{\partial z_u^{'}}
    = \frac{\partial \Lapl_{ssl}^{user}(u)}{\partial s_u^{'}} \frac{\partial s_u^{'}}{\partial z_u^{'}} 
\end{gather}
where,
\begin{align}
    \frac{\partial s_u^{'}}{\partial z_u^{'}}
    &=\frac{\partial}{\partial z_u^{'}} \left( \frac{z_u^{'}}{\left \| z_u^{'} \right \|} \right)\\ \nonumber
    &= \frac{1}{\left \| z_u^{'} \right \|}\mathrm{I}  + z_u^{'}\left( \frac{\partial \left(1/\left \| z_u^{'} \right \|\right)}{\partial z_u^{'}} \right) \\ \nonumber
    &= \frac{1}{\left \| z_u^{'} \right \|} \left( \mathrm{I} - \frac{z_u^{'} {z_u^{'}}^T}{{\left \| z_u^{'} \right \|}^2} \right) \\ \nonumber
    &= \frac{1}{\left \| z_u^{'} \right \|} \left( \mathrm{I} - s_u^{'} {s_u^{'}}^T \right)
\end{align}

\begin{align}
    &\frac{\partial \Lapl_{ssl}^{user}(u)}{\partial s_u^{'}}
    = -\frac{\partial}{\partial s_u^{'}} \left( {s_u^{'}}^T s_u^{''} / \tau \right) + \frac{\partial}{\partial s_u^{'}} \log \sum_{v\in \mathcal{U}} \exp \left( {s_u^{'}}^T s_v^{''} /\tau \right) \\ \nonumber
    =& \frac{1}{\tau} \left( \frac{\sum\limits_{v\in\mathcal{U}} {s_v^{''}}^T \exp\left( {s_u^{'}}^T s_v^{''} /\tau \right)}{\sum\limits_{v \in \mathcal{U}} \exp \left( {s_u^{'}}^T s_v^{''} /\tau \right)}  - {s_u^{''}}^T\right) \\ \nonumber
    =& \frac{1}{\tau} \left( \sum\limits_{v \in \mathcal{U} \setminus \{u\}} {s_v^{''}}^T \frac{\exp(s_u^{'} s_v^{''}/\tau)}{\sum\limits_{v\in \mathcal{U}} \exp({s_u^{'}}^T s_v^{''} / \tau)} + {s_u^{''}}^T \frac{\exp(s_u^{'} s_u^{''}/\tau)}{\sum\limits_{v\in \mathcal{U}} \exp({s_u^{'}}^T s_v^{''} / \tau)} - {s_u^{''}}^T \right) \\ \nonumber
    =& \frac{1}{\tau} \left( \sum\limits_{v \in \mathcal{U} \setminus \{u\}} {s_v^{''}}^TP_{uv} + {s_u^{''}}^T \left( P_{uu} - 1 \right) \right)
\end{align}
where,
\begin{align}
    P_{uv} &= \frac{\exp({s_u^{'}}^T s_v^{''}/\tau)}{\sum\limits_{v\in \mathcal{U}} \exp({s_u^{'}}^T s_v^{''} / \tau)}\\
    P_{uu} &= \frac{\exp({s_u^{'}}^T s_u^{''}/\tau)}{\sum\limits_{v\in \mathcal{U}} \exp({s_u^{'}}^T s_v^{''} / \tau)}
\end{align}
$P_{uv}$ is the likelihood for $s_v^{''}$ \wrt all samples within $\mathcal{U}$.
Therefore, we have
\begin{align}
    &\frac{\partial \Lapl_{ssl}^{user}(u)}{\partial z_u^{'}}
    = \frac{1}{\tau \left \| z_u^{'} \right \|} \left\{ \sum\limits_{v \in \mathcal{U} \setminus \{u\}} {s_v^{''}}^TP_{uv} + {s_u^{''}}^T \left( P_{uu} - 1 \right) \right\} \left( \mathrm{I} - s_u^{'}{s_u^{'}}^T \right) \\ \nonumber
    =& \frac{1}{\tau \left \| z_u^{'} \right \|} \left\{ \begin{matrix} \underbrace{ \left( s_u^{''} - ({s_u^{'}}^Ts_u^{''}) s_u^{'} \right)^T \left( P_{uu} - 1 \right) } \\ c(u)\end{matrix} + \sum\limits_{v \in \mathcal{U} \setminus \{u\}}  \begin{matrix} \underbrace{ \left( s_v^{''} - ({s_u^{'}}^Ts_v^{''})s_u^{'} \right)^T P_{uv} } \\ c(v)\end{matrix} \right\}
\end{align}

\end{document}